# How semiconductor nanoplatelets form

Andreas Riedinger[1,†], Florian D. Ott[1,†], Aniket Mule[1], Sergio Mazzotti[1], Philippe N. Knüsel[1], Stephan J. P. Kress[1], Ferry Prins[1], Steven C. Erwin[2]* and David J. Norris[1]*

[1]Optical Materials Engineering Laboratory, Department of Mechanical and Process Engineering, ETH Zurich, 8092 Zurich, Switzerland.
[2]Center for Computational Materials Science, Naval Research Laboratory, Washington, D.C. 20375 USA.

[†]These authors contributed equally to this work.
*email: steve.erwin@nrl.navy.mil; dnorris@ethz.ch

**ABSTRACT**

Colloidal nanoplatelets—quasi-two-dimensional sheets of semiconductor exhibiting efficient, spectrally pure fluorescence—form when liquid-phase syntheses of spherical quantum dots are modified. Despite intense interest in their properties, the mechanism behind their anisotropic shape and precise atomic-scale thickness remains unclear, and even counterintuitive when their crystal structure is isotropic. One commonly accepted explanation is that nanoclusters nucleate within molecular templates and then fuse. Here, we test this mechanism for zincblende nanoplatelets and show that they form instead due to an intrinsic instability in growth kinetics. We synthesize CdSe and $CdS_{1-x}Se_x$ nanoplatelets in template- and solvent-free isotropic melts containing only cadmium carboxylate and chalcogen, a finding incompatible with previous explanations. Our model, based on theoretical results showing enhanced growth on narrow surface facets, rationalizes nanoplatelet formation and experimental dependencies on temperature, time, and carboxylate length. Such understanding should lead to improved syntheses, controlled growth on surfaces, and broader libraries of nanoplatelet materials.

Colloidal quantum dots are fluorescent semiconductor nanocrystals nearly spherical in shape. Produced via liquid-phase chemical syntheses[1], their protocols have been manipulated to explore other shapes such as rods[2], tetrapods[3], and arrows[4,5]. For materials with an anisotropic crystal lattice, such forms arise naturally when growth is induced along specific crystallographic directions[2]. For example, rods result when growth along the *c*-axis of hexagonal (wurtzite) CdSe is enhanced by selective binding of surface ligands[4] or by kinetically over-driving the reaction[2]. Even for isotropic crystal structures, such as cubic (zincblende) CdSe, non-spherical shapes can be obtained using ligands with temperature-dependent binding[6]. This allows control over addition to {001} and {111} facets, leading to cubes, tetrahedrons, or branched particles, depending on temperature.

However, while the formation of the above shapes is understood, the mechanism behind another increasingly important nanostructure is less clear. A variety of quasi-two-dimensional (2D) semiconductor particles known as nanoribbons[7], nanoplatelets (NPLs)[8], nanosheets[9], and quantum belts[10] have recently been reported. Among these, CdSe NPLs have been the most heavily studied. Such samples contain zincblende platelets that vary in lateral dimensions but are remarkably uniform in thickness[8], controllable from 3 to 7 monolayers[11]. Because their optoelectronic properties are governed by this precise atomic-scale thickness, NPLs provide unique behavior[11-15], such as spectrally pure fluorescence[16], large absorption cross-sections[17], enhanced energy transfer[14], and boosted optical gain[18]. These characteristics are useful for light-emitting devices[19], field-effect transistors[20], solar cells[14], and lasers[18].

Despite intense interest in these properties, the formation of zincblende CdSe NPLs remains puzzling[21], especially as their highly anisotropic shape arises from an isotropic cubic crystal structure with primarily {001} facets exposed[22]. Consequently, their shape cannot be explained by differential growth on distinct crystallographic facets. Rather, another mechanism must be responsible. One possibility, proposed to explain wurtzite

CdSe platelets, is that growth occurs within soft molecular templates[23]. For example, Cd precursors of the type $CdX_2(R-NH_2)_2$ can produce lamellar mesophases[24]. When such phases are treated with chalcogens near room temperature, "magic-sized" clusters [*e.g.* $(CdSe)_{13}$] can form within this template and organize into stacked rectangular arrays[10]. With mild annealing, these arrays fuse via oriented attachment[25-27], yielding wurtzite CdSe platelets[28] that are dispersible via exfoliation[7,10,24]. Mesophases have also been implicated in the growth of PbS nanosheets[9], which form at 100 °C.

Here we examine whether zincblende CdSe NPLs form via a similar mechanism. We find no evidence for this explanation and conclude that these NPLs form instead due to an intrinsic instability in their shape during growth. This previously unknown mechanism can lead to new and improved NPL processes and materials as well as a broader understanding of nanostructure shape control.

**Results**

**Role of carboxylates in NPL syntheses.** Standard liquid-phase protocols[8,29] for zincblende CdSe NPLs involve a long-chain $Cd(carboxylate)_2$ [either $Cd(myristate)_2$ or $Cd(oleate)_2$], which is heated with Se powder in a non-coordinating solvent (1-octadecene, ODE). Between 180 and 240 °C, a short-chain $Cd(carboxylate)_2$ [*e.g.* $Cd(acetate)_2$] is added, and then the mixture is maintained at 240 °C for 5-15 min. The addition of this short-chain carboxylate appears to be the critical step. The NPL thickness is regulated by the addition temperature and the NPL lateral size by the subsequent reaction time[21]. Without this addition, quantum dots and other shapes form instead of NPLs.

The apparent importance of the short-chain carboxylate provides a clue to the growth mechanism. To reveal this, we added short-chain carboxylic acids of various lengths to the standard NPL protocol (see Supplementary Information). We found that NPL formation depended on both the carbon-chain length ($C_x$) and concentration. Acetic anhydride, which decomposes to acetic acid ($C_2$), yielded NPLs at all tested concentrations, propionic acid ($C_3$) only at medium to high concentrations, and butyric acid ($C_4$) at none (producing primarily quantum dots). Thus, shorter carboxylic acids lead to easier formation of NPLs in the standard liquid-phase synthesis.

We could explain this by repeating these experiments without Se. In this case, NPLs could not form, but the intermediates produced in the standard protocol when a short-chain carboxylic acid is added to long-chain Cd(oleate)$_2$ could be identified. We injected two equivalents of $C_2$, $C_3$, or $C_4$ at 180 °C into a transparent solution of ODE containing Cd(oleate)$_2$. For $C_2$ and $C_3$, the solutions immediately turned cloudy, but not for $C_4$. After 30 min, the products were isolated, yielding Cd(acetate)$_2$, Cd(oleate)$_{2-x}$(propionate)$_x$, and nearly pure Cd(oleate)$_2$, for $C_2$, $C_3$, and $C_4$, respectively (see Supplementary Information). This indicates that shorter carboxylic acids are more likely to undergo ligand exchange with Cd(oleate)$_2$ to form mixed carboxylates of the type Cd(long carboxylate)$_{2-x}$(short carboxylate)$_x$. The replacement of long with short chains lowers the solubility of the carboxylate in ODE at the NPL reaction temperature, leading to the observed cloudiness for $C_2$ and $C_3$. Therefore, we hypothesize that the addition of short-chain Cd(acetate)$_2$ during the standard NPL protocol lowers the solubility of the Cd precursor by forming mixed carboxylates, which then phase separate as droplets. The solubility of the Cd(carboxylate)$_2$ in the solvent then determines if NPLs or quantum dots grow. If the precursor dissolves, quantum dots result; if the precursor is insoluble, NPLs form.

Indeed, when we heated Se with only long-chain Cd(myristate)$_2$ in ODE to 200 °C, the Cd(myristate)$_2$ dissolved at ~100 °C and quantum dots were obtained (Fig. 1a). With only short-chain Cd(propionate)$_2$ instead, liquid droplets phase separated in the reaction solution above 180 °C due to the insolubility of this precursor in ODE. CdSe NPLs formed within these droplets, which solidified upon cooling. The resulting NPLs were 4-monolayers thick as indicated by optical absorption features at 437 and 461 nm (Fig. 1a)[22].

These results indicate that a mixture of short- and long-chained carboxylates is actually not required to obtain NPLs if the Cd precursor is already insoluble under the reaction conditions. Furthermore, because NPLs grew in the phase-separated droplets of short-chain Cd(propionate)$_2$ rather than in solution, even the solvent is apparently unnecessary.

To test this, we mixed either long-chain Cd(myristate)$_2$ or short-chain Cd(propionate)$_2$ with Se powder without solvent and heated the blends at 200 °C for 18 h under N$_2$. Above the melting point of the carboxylate [~100 °C for Cd(myristate)$_2$ and ~180 °C for Cd(propionate)$_2$] a black, viscous slurry formed, which slowly turned yellow. Both reactions yielded 4-monolayer CdSe NPLs[22] (Fig. 1b), but with different lateral sizes. Thus, NPLs can form in a melt of a single Cd(carboxylate)$_2$, either long- or short-chained, with the length only modifying growth kinetics. We conclude that the addition of short-chain carboxylates is only critical in the standard liquid-phase protocols[8,29]. There, short carboxylates are required to phase separate the Cd precursor from the solvent (ODE). Thus, even in the standard protocols, NPLs form in "solvent-free" environments. When we removed the solvent we could grow NPLs with either long or short carboxylates.

**Absence of molecular mesophases and oriented attachment.** In solvent-free environments, it is unlikely that lamellar mesophases are present to template NPL growth. For short-chain Cd(propionate)$_2$ such phases are not even known[30]. Nevertheless, we looked for mesophases in the reaction melt by performing *in situ* wide-angle X-ray scattering (WAXS) (Fig. 1c). For the short-chain Cd(propionate)$_2$ we observed a polymorphic transition around 100 °C and then above the melting point (180 °C) only a broad peak, indicative of a disordered amorphous phase. Upon cooling back to 25 °C, the system did not recrystallize and this broad peak remained, as Cd(propionate)$_2$ was frozen in a glassy state. This order-to-disorder transition was further supported by differential scanning calorimetry (DSC) measurements (see Supplementary Information). Attempts to observe mesophases in Cd(propionate)$_2$ melts (also in the presence of Se) via polarized optical microscopy were also negative (see Supplementary Information).

Thus, our findings indicate that zincblende CdSe NPLs can form in completely isotropic environments [*e.g.*, Cd(propionate)$_2$ above 180 °C]. Spatial constraints from molecular mesophases are not necessary to obtain their highly anisotropic shapes. In the case of PbS nanosheets[9], molecular templates were invoked to explain how pre-formed nanoclusters of PbS fuse via 2D oriented attachment[23]. Even if templates are not involved in the growth of zincblende CdSe NPLs, oriented attachment could still play a role. Nanoclusters would need to be present and somehow guided to form sheets. However, as we now describe, our results are inconsistent with this mechanism.

Because NPLs exhibit precise atomic-scale thicknesses, the nanoclusters would need to be extremely small and uniform in size. These requirements place the clusters in the "magic-size" regime[31,32] [*e.g.* (CdSe)$_{13}$, (CdSe)$_{34}$, and Cd$_{34}$S$_{14}$]. However, such clusters have never been observed during syntheses of zincblende NPLs[23]. This is consistent with their poor stability (versus larger quantum dots)[28]. Nevertheless, to examine their possible involvement, we synthesized CdS$_{1-x}$Se$_x$ NPLs. Although such NPLs could arise from alloyed magic-sized clusters of the type CdS$_{1-x}$Se$_x$, these species are not known, presumably because they are even less stable than CdSe clusters. If they did exist, they would occur only for specific stoichiometries, and their composition would not be easily varied. Hence, if a cluster-oriented-attachment mechanism were operable, homogeneous CdS$_{1-x}$Se$_x$ NPLs with tunable composition would not be expected when cadmium carboxylates were heated with different Se:S ratios.

Our experiments yielded just such homogeneous CdS$_{1-x}$Se$_x$ NPLs. We observed monotonic shifts of the optical features with increasing Se content (Fig. 2a), consistent with alloying. The NPLs (Fig. 2b) also contained a completely homogeneous distribution of Se and S according to energy dispersive X-ray analyses (Fig. 2c). No trace of mixed-phase

NPLs was detected. Compared to pure CdSe or CdS NPLs, these samples also exhibited broadened absorption features and larger Stokes shifts, similar to trends observed for alloyed $CdS_{1-x}Se_x$ quantum dots[33]. Thus, based on all of our data, we conclude that neither molecular templates nor oriented attachment is involved in the formation of zincblende CdSe NPLs.

**Chalcogenide precursor chemistry.** Before presenting an alternative explanation, we briefly address the chalcogenide precursor. As already shown, we can obtain CdSe, CdS, or $CdS_{1-x}Se_x$ NPLs in solvent-free melts of $Cd(propionate)_2$ with elemental Se, S, or their mixtures. The same procedure with Te, however, did not yield CdTe NPLs. This implies that a specific chalcogen-to-chalcogenide reduction pathway exists for Se and S but not Te. Because Cd is already in its highest oxidation state, Se and S must be reduced by the only other possibility, the propionate. To identify the pathway, we collected thermogravimetric (TGA) and DSC data while measuring electron-ionization mass spectroscopy (MS) of volatile species. We analyzed four samples: $Cd(propionate)_2$ alone and with equimolar amounts of either S, Se, or Te. In all cases, weight loss below 180 °C was negligible. Above this temperature, differences between group 1 [$Cd(propionate)_2$ alone or with Te] and group 2 [$Cd(propionate)_2$ with Se or S] were observed (Fig. 3a,b). While both groups showed melting of $Cd(propionate)_2$ at 180 °C, group 2 also showed melting of elemental Se (220 °C) and S (115 °C). Group 1 exhibited exothermic decomposition at 227 °C, which was absent in group 2.

These findings can be combined with the simultaneously recorded MS data. In general, MS of transition-metal propionates shows two distinct decomposition products: propionyl radicals, observed as propionyl cations with a mass-to-charge (m/z) ratio of 57, and 2-pentanone with m/z of 86[34,35]. In group 1, both species were observed below 250 °C, but not in group 2 (Fig. 3c and Supplementary Information). This leads to our proposed reaction mechanism for CdSe (or CdS) NPLs (Fig. 3e). The absence of propionyl radical in the presence of Se and S strongly suggests the formation of dipropionyl selenide and sulfide. These molecules can react with Cd(propionate)$_2$ to yield CdSe or CdS, releasing propionate anions and propionyl cations. The latter decomposes to carbon monoxide, ethane, and protons (Fig. 3e). These protons can then be captured by free propionate to yield propionic acid. Indeed, traces of this molecule were detected, but only for Se and S (Fig. 3d). Thus, dipropionyl selenide and sulfide are apparently the reactive chalcogenide precursors in NPL growth.

Beyond NPLs, such precursors could be exploited in nanocrystal synthesis, where elemental Se or S is often combined with metal carboxylates. Recently, organosulfide precursors with tailored reactivity were used to improve sulfur-based quantum dots[36]. Libraries of diacyl and diaryl selenide precursors could similarly allow controlled reactivity.

**Theoretical model of NPL growth.** We now present a model that accounts for the highly anisotropic shape of zincblende NPLs as well as other key experimental observations. Namely, when such NPLs grow laterally[16,37,38], their thickness remains largely fixed. Thickness can, however, increase at long times and high temperatures, *e.g.* as seen for CdS[39] and CdTe[40]. During our experiments, we also observed transitions from 3 to 4 monolayers. Thus, our model must also address why the optical features of the thinner NPLs disappear while those of the thicker NPLs emerge (Fig. 2d).

To explain these observations, we applied the general theory of 2D nucleation and growth[41] to CdSe nanocrystals passivated by Cd-acetate ligands. Growth occurs when new islands nucleate on the exposed facets of the nanocrystals. Once an island reaches a critical size, it becomes stable and has a thermodynamic driving force to grow[41,42]. Because the precursor melt is highly concentrated, diffusion of material to the growing island will be fast. Thus, the overall growth mode will be governed by island nucleation. On large planar surfaces this nucleation process has a fixed activation barrier. Below we show that on narrow facets with dimensions below the critical island size, the nucleation barrier can decrease. This explains why a cubic crystal structure can exhibit anisotropic growth. Faster growth can occur on thin facets compared to large planar surfaces.

Our model assumes an existing zincblende CdSe nanocrystal bounded by Cd-terminated {001} facets. We then investigated how the energy changes, for different facet dimensions, when an island nucleates and grows to eventually cover the facet. For simplicity we also assumed that Cd and Se atoms are always incorporated together, as a "monomer." The change in total energy of the system can be expressed approximately as $\Delta E = \Delta V\, E_V + \Delta A\, E_A + \Delta L\, E_L$. Here $\Delta V$, $\Delta A$, and $\Delta L$ are the change in volume, area, and edge length of the crystallite due to growth, while $E_V$, $E_A$, and $E_L$ are the respective energies per unit volume, area, and length. Energetically, solid CdSe is more stable than the melt and thus $E_V$ is negative, while $E_A$ and $E_L$ are positive. Hence, the change in total energy when new material is added will be minimized when $\Delta A$ and $\Delta L$ are minimized.

We evaluated $\Delta E$ in two regimes: wide and narrow facets (see Fig. 4a). On wide facets, the island nucleates most easily in a corner. We assumed for simplicity that it then grows as a square with sides parallel to the crystallite facets. Its energy is given by $E^{\text{wide}}(a) = (L_1\, E_V)\, a + (4L_1 E_A + 2E_L)\, a^{1/2} + 4L_1 E_L$ where $L_1$ is the height of one monolayer and $a$ is the island area. For narrow facets, the island minimizes its energy by instead maintaining a single, short step edge spanning the facet. In this case, once the island is large enough to span the facet, no additional edge energy needs to be paid as the layer is completed. Hence, in this regime the energy is given by a family of linear relationships, $E_m^{\text{narrow}}(a) = (L_1\, E_V + 2\, E_A/m)\, a + 2m L_1^2 E_A + (m+4)L_1 E_L$, where $m$ is the facet thickness in monolayers.

To quantify the above relationships, we used density-functional theory (DFT) to obtain $E_A = 5.7$ meV/Å² for acetate-passivated CdSe(001) under Se-rich conditions and $E_L = 37.1$ meV/Å for monolayer steps on CdSe(001) (see Supplementary Information). Calculating $E_V$ from first principles is more challenging because the reference state is a melt. Instead, we estimated $E_V$ (−1.5 meV/Å³) from the experimental absence of 2-monolayer-thick NPLs (see Supplementary Information).

Figure 4b shows the resulting island energies in the two regimes. $E^{wide}(a)$ (black curve) has the classic shape from standard nucleation theory while $E_m^{narrow}(a)$ behaves very differently. The nucleation barriers on narrow facets (denoted by the colored points) are smaller than on wide facets. For example, the barrier on 4-monolayer facets is reduced from 1.6 to 1.3 eV. This implies much faster growth on narrow facets—three orders of magnitude faster at the growth temperature of 200 °C. Even small changes in the facet thickness dramatically affect the growth rate. One additional layer slows the growth by a factor ten. Consequently, even cubical seed crystallites can evolve into flat NPLs due to random fluctuations. This instability can lead to anisotropic NPLs even for materials with a cubic crystal structure. To verify this claim statistically, we used the kinetic Monte Carlo (kMC) method to simulate the atomistic growth of seed crystallites at finite temperatures according to our growth model. We indeed observed that a large fraction of crystallites, including those with initially cubical shapes, evolved into highly anisotropic NPLs (see Discussion, Movies S1 and S2, and the Supplementary Information).

Because we constrained our value of $E_V$ to match the experimental absence of 2-monolayer NPLs, the island energy in Fig. 4b for $m = 2$ increases indefinitely with size. For thicker (*e.g.* 4-, 5-, and 6-monolayer) platelets, growth is suppressed with thickness as their nucleation barriers approach the wide-facet limit. However, for sufficiently large island areas, these thicker facets eventually become more stable, as indicated by their increasingly negative slope beyond the barrier. Thus, the model predicts that the rapid growth of thin, 3-monolayer NPLs will eventually yield, at high temperature and long reaction times, to slower growth of thicker platelets—in agreement with experiments.

These trends can be quantified using a system of coupled rate equations describing how a mixture of CdSe monomers and NPLs of different thicknesses evolves in time. We estimated nucleation rates from the barriers in Fig. 4b and equilibrium constants from the reaction energies of NPLs of a given thickness. Figure 4c shows the time evolution of the distribution of monomers within an ensemble of NPLs at 200 °C. At first, only 3-monolayer NPLs appear, but then 4-, 5-, and 6-monolayer NPLs appear in succession. Because the time needed for each new thickness increases by several orders of magnitude, the NPL thickness can be accurately controlled with time and temperature. Using the parameters from Fig. 4b, this simple model is consistent with our experimental observation of a transition from 3- to 4-monolayer platelets, as shown in Fig. 2d,e. Moreover, it suggests that 3-monolayer NPLs are dissolving while new 4-monolayer NPLs are forming.

## Discussion

Our model also reveals which material properties promote platelet growth. For example, the thermodynamic driving force for adding material to a facet is maximized when $|L_1 E_V|$ is large and $2E_A/m$ is small. The first term is fixed by the properties of the bulk material and growth solution at a given temperature. Thus, a simple strategy to obtain thin NPLs (with small $m$) is to keep $E_A$ small. The growth of NPLs is then favored by using strongly binding ligands. Conversely, weak surfactants will lead to thicker platelets.

We also find that the activation barrier for island nucleation scales inversely with $|E_V|$ and quadratically with $E_A$ and $E_L$. Hence, if $|E_V|$ is sufficiently large, the barrier can be easily overcome on any facet, and the instability leading to NPLs disappears. On the other hand, if $|E_V|$ is small, or if $E_A$ and $E_L$ are large, the barrier may be so high that it is insurmountable under experimental conditions. In that case no growth will occur. Further, the instability toward NPLs only exists if $E_L$ is positive. Even if $E_L$ is very small or zero, the nucleation barrier on narrow facets will be similar to that on wide facets. Thus, NPL growth requires a careful balance of $E_V$, $E_A$, and $E_L$. The evaluation of these parameters for different material-ligand combinations should allow a systematic investigation of which materials are suitable for growing NPLs.

We have made preliminary efforts in this direction by attempting to synthesize new NPL materials (see Supplementary Information). For example, we heated mercury acetate with Se. We saw no evidence of HgSe NPL formation; rather, large irregular crystals were obtained. DFT predicts $E_A$ = 2.5 meV/Å$^2$ and $E_L$ = 9.3 meV/Å for HgSe, which leads to a wide-facet barrier of only 0.25 eV, assuming the same $E_V$ as used in CdSe. This barrier is negligible at the growth temperature and consequently no NPL instability is expected.

Interestingly, atomically flat CsPbBr$_3$ perovskite nanoplatelets with a cubic crystal structure have also been synthesized recently[43]. Their thickness was controlled by the reaction temperature. Below a critical temperature, no CsPbBr$_3$ NPLs were formed. With increasing temperature, increasingly thick NPLs (one to five monolayers) were observed. Above a critical temperature, CsPbBr$_3$ with cubic crystal habits were obtained[44]. These findings can be easily rationalized by our model, demonstrating the potential of our results for developing new NPL materials guided by a rational theoretical model.

## Methods

Details related to synthetic methods, characterization, calculations, and simulations are presented in the Supplementary Information.

## References

1. Murray, C. B., Norris, D. J. & Bawendi, M. G. Synthesis and characterization of nearly monodisperse CdE (E = sulfur, selenium, tellurium) semiconductor nanocrystallites. *J. Am. Chem. Soc.* **115**, 8706-8715 (1993).
2. Peng, X., Manna, L., Yang, W., Wickham, J., Scher, E., Kadavanich, A. & Alivisatos, A. P. Shape control of CdSe nanocrystals. *Nature* **404**, 59-61 (2000).

3. Manna, L., Milliron, D. J., Meisel, A., Scher, E. C. & Alivisatos, A. P. Controlled growth of tetrapod-branched inorganic nanocrystals. *Nat. Mater.* **2**, 382-385 (2003).
4. Manna, L., Scher, E. C. & Alivisatos, A. P. Synthesis of soluble and processable rod-, arrow-, and teardrop-, and tetrapod-shaped CdSe nanocrystals. *J. Am. Chem. Soc.* **122**, 12700-12706 (2000).
5. Milliron, D. J., Hughes, S. M., Cui, Y., Manna, L., Li, J., Wang, L.-W. & Alivisatos, A. P. Colloidal nanocrystal heterostructures with linear and branched topology. *Nature* **430**, 190-195 (2004).
6. Liu, L., Zhuang, Z., Xie, T., Wang, Y.-G., Li, J., Peng, Q. & Li, Y. Shape control of CdSe nanocrystals with zinc blende structure. *J. Am. Chem. Soc.* **131**, 16423-16429 (2009).
7. Joo, J., Son, J. S., Kwon, S. G., Yu, J. H. & Hyeon, T. Low-temperature solution-phase synthesis of quantum well structured CdSe nanoribbons. *J. Am. Chem. Soc.* **128**, 5632-5633 (2006).
8. Ithurria, S. & Dubertret, B. Quasi 2D colloidal CdSe platelets with thicknesses controlled at the atomic level. *J. Am. Chem. Soc.* **130**, 16504-16505 (2008).
9. Schliehe, C., Juarez, B. H., Pelletier, M., Jander, S., Greshnykh, D., Nagel, M., Meyer, A., Foerster, S., Kornowski, A., Klinke, C. & Weller, H. Ultrathin PbS sheets by two-dimensional oriented attachment. *Science* **329**, 550-553 (2010).
10. Liu, Y.-H., Wang, F., Wang, Y., Gibbons, P. C. & Buhro, W. E. Lamellar assembly of cadmium selenide nanoclusters into quantum belts. *J. Am. Chem. Soc.* **133**, 17005-17013 (2011).
11. Ithurria, S., Tessier, M. D., Mahler, B., Lobo, R., Dubertret, B. & Efros, A. Colloidal nanoplatelets with two-dimensional electronic structure. *Nat. Mater.* **10**, 936-941 (2011).
12. Achtstein, A. W., Schliwa, A., Prudnikau, A., Hardzei, M., Artemyev, M. V., Thomsen, C. & Woggon, U. Electronic structure and exciton–phonon interaction in two-dimensional colloidal CdSe nanosheets. *Nano Lett.* **12**, 3151-3157 (2012).
13. Pelton, M., Ithurria, S., Schaller, R. D., Dolzhnikov, D. S. & Talapin, D. V. Carrier cooling in colloidal quantum wells. *Nano Lett.* **12**, 6158-6163 (2012).
14. Rowland, C. E., Fedin, I., Zhang, H., Gray, S. K., Govorov, A. O., Talapin, D. V. & Schaller, R. D. Picosecond energy transfer and multiexciton transfer outpaces Auger recombination in binary CdSe nanoplatelet solids. *Nat. Mater.* **14**, 484-489 (2015).
15. Guzelturk, B., Erdem, O., Olutas, M., Kelestemur, Y. & Demir, H. V. Stacking in colloidal nanoplatelets: Tuning excitonic properties. *ACS Nano* **8**, 12524-12533 (2014).
16. Olutas, M., Guzelturk, B., Kelestemur, Y., Yeltik, A., Delikanli, S. & Demir, H. V. Lateral size-dependent spontaneous and stimulated emission properties in colloidal CdSe nanoplatelets. *ACS Nano* **9**, 5041-5050 (2015).
17. Yeltik, A., Delikanli, S., Olutas, M., Kelestemur, Y., Guzelturk, B. & Demir, H. V. Experimental determination of the absorption cross-section and molar extinction coefficient of colloidal CdSe nanoplatelets. *J. Phys. Chem. C* **119**, 26768-26775 (2015).
18. She, C., Fedin, I., Dolzhnikov, D. S., Demortière, A., Schaller, R. D., Pelton, M. & Talapin, D. V. Low-threshold stimulated emission using colloidal quantum wells. *Nano Lett.* **14**, 2772-2777 (2014).
19. Chen, Z., Nadal, B., Mahler, B., Aubin, H. & Dubertret, B. Quasi-2D colloidal semiconductor nanoplatelets for narrow electroluminescence. *Adv. Funct. Mater.* **24**, 295-302 (2014).
20. Lhuillier, E., Pedetti, S., Ithurria, S., Heuclin, H., Nadal, B., Robin, A., Patriarche, G., Lequeux, N. & Dubertret, B. Electrolyte-gated field effect transistor to probe the surface defects and morphology in films of thick CdSe colloidal nanoplatelets. *ACS Nano* **8**, 3813-3820 (2014).
21. Ithurria, S., Bousquet, G. & Dubertret, B. Continuous transition from 3D to 1D confinement observed during the formation of CdSe nanoplatelets. *J. Am. Chem. Soc.* **133**, 3070-3077 (2011).
22. Hutter, E. M., Bladt, E., Goris, B., Pietra, F., van der Bok, J. C., Boneschanscher, M. P., de Mello Donegá, C., Bals, S. & Vanmaekelbergh, D. Conformal and atomic characterization of ultrathin CdSe platelets with a helical shape. *Nano Lett.* **14**, 6257-6262 (2014).
23. Wang, F., Wang, Y., Liu, Y.-H., Morrison, P. J., Loomis, R. A. & Buhro, W. E. Two-dimensional semiconductor nanocrystals: Properties, templated formation, and magic-size nanocluster intermediates. *Acc. Chem. Res.* **48**, 13-21 (2015).

24. Son, J. S., Wen, X.-D., Joo, J., Chae, J., Baek, S.-i., Park, K., Kim, J. H., An, K., Yu, J. H., Kwon, S. G., Choi, S.-H., Wang, Z., Kim, Y.-W., Kuk, Y., Hoffmann, R. & Hyeon, T. Large-scale soft colloidal template synthesis of 1.4 nm thick CdSe nanosheets. *Angew. Chem. Int. Ed.* **48**, 6861-6864 (2009).
25. Penn, R. L. & Banfield, J. F. Imperfect oriented attachment: Dislocation generation in defect-free nanocrystals. *Science* **281**, 969-971 (1998).
26. Yu, J. H., Joo, J., Park, H. M., Baik, S.-I., Kim, Y. W., Kim, S. C. & Hyeon, T. Synthesis of quantum-sized cubic ZnS nanorods by the oriented attachment mechanism. *J. Am. Chem. Soc.* **127**, 5662-5670 (2005).
27. Cho, K.-S., Talapin, D. V., Gaschler, W. & Murray, C. B. Designing PbSe nanowires and nanorings through oriented attachment of nanoparticles. *J. Am. Chem. Soc.* **127**, 7140-7147 (2005).
28. Wang, Y., Zhang, Y., Wang, F., Giblin, D. E., Hoy, J., Rohrs, H. W., Loomis, R. A. & Buhro, W. E. The magic-size nanocluster $(CdSe)_{34}$ as a low-temperature nucleant for cadmium selenide nanocrystals: Room-temperature growth of crystalline quantum platelets. *Chem. Mater.* **26**, 2233-2243 (2014).
29. Mahler, B., Nadal, B., Bouet, C., Patriarche, G. & Dubertret, B. Core/shell colloidal semiconductor nanoplatelets. *J. Am. Chem. Soc.* **134**, 18591-18598 (2012).
30. Mirnaya, T. A. & Volkov, S. V. in *Green Industrial Applications of Ionic Liquids*, edited by R. D. Rogers, K. R. Seddon, and S. V. Volkov (Springer Netherlands, Dordrecht, 2002), pp. 439-456.
31. Del Ben, M., Havenith, R. W. A., Broer, R. & Stener, M. Density functional study on the morphology and photoabsorption of CdSe nanoclusters. *J. Phys. Chem. C* **115**, 16782-16796 (2011).
32. Herron, N., Calabrese, J. C., Farneth, W. E. & Wang, Y. Crystal structure and optical properties of $Cd_{32}S_{14}(SC_6H_5)_{36} \cdot DMF_4$, a cluster with a 15 angstrom CdS core. *Science* **259**, 1426-1428 (1993).
33. Garrett, M. D., Dukes, A. D., McBride, J. R., Smith, N. J., Pennycook, S. J. & Rosenthal, S. J. Band edge recombination in CdSe, CdS and $CdS_xSe_{1-x}$ alloy nanocrystals observed by ultrafast fluorescence upconversion: The effect of surface trap states. *J. Phys. Chem. C* **112**, 12736-12746 (2008).
34. Nasui, M., Mos, R. B., Petrisor Jr, T., Gabor, M. S., Varga, R. A., Ciontea, L. & Petrisor, T. Synthesis, crystal structure and thermal decomposition of a new copper propionate $[Cu(CH_3CH_2COO)_2] \cdot 2H_2O$. *J. Anal. Appl. Pyrol.* **92**, 439-444 (2011).
35. Kercher, J. P., Fogleman, E. A., Koizumi, H., Sztáray, B. & Baer, T. Heats of formation of the propionyl ion and radical and 2,3-pentanedione by threshold photoelectron photoion coincidence spectroscopy. *J. Phys. Chem. A* **109**, 939-946 (2005).
36. Hendricks, M. P., Campos, M. P., Cleveland, G. T., Jen-La Plante, I. & Owen, J. S. A tunable library of substituted thiourea precursors to metal sulfide nanocrystals. *Science* **348**, 1226-1230 (2015).
37. Bouet, C., Mahler, B., Nadal, B., Abecassis, B., Tessier, M. D., Ithurria, S., Xu, X. & Dubertret, B. Two-dimensional growth of CdSe nanocrystals, from nanoplatelets to nanosheets. *Chem. Mater.* **25**, 639-645 (2013).
38. Tessier, M. D., Spinicelli, P., Dupont, D., Patriarche, G., Ithurria, S. & Dubertret, B. Efficient exciton concentrators built from colloidal core/crown CdSe/CdS semiconductor nanoplatelets. *Nano Lett.* **14**, 207-213 (2013).
39. Li, Z., Qin, H., Guzun, D., Benamara, M., Salamo, G. & Peng, X. Uniform thickness and colloidal-stable CdS quantum disks with tunable thickness: Synthesis and properties. *Nano Res.* **5**, 337-351 (2012).
40. Pedetti, S., Nadal, B., Lhuillier, E., Mahler, B., Bouet, C., Abécassis, B., Xu, X. & Dubertret, B. Optimized synthesis of CdTe nanoplatelets and photoresponse of CdTe nanoplatelets films. *Chem. Mater.* **25**, 2455-2462 (2013).
41. Lovette, M. A., Browning, A. R., Griffin, D. W., Sizemore, J. P., Snyder, R. C. & Doherty, M. F. Crystal shape engineering. *Ind. Eng. Chem. Res.* **47**, 9821-9833 (2008).
42. Ohara, M. & Reid, R. C., *Modeling Crystal Growth Rates from Solution*. (Prentice-Hall, 1973, 1973).
43. Bekenstein, Y., Koscher, B. A., Eaton, S. W., Yang, P. & Alivisatos, A. P. Highly luminescent colloidal nanoplates of perovskite cesium lead halide and their oriented assemblies. *J. Am. Chem. Soc.* **137**, 16008-16011 (2015).
44. Protesescu, L., Yakunin, S., Bodnarchuk, M. I., Krieg, F., Caputo, R., Hendon, C. H., Yang, R. X., Walsh, A. & Kovalenko, M. V. Nanocrystals of cesium lead halide perovskites ($CsPbX_3$, X = Cl, Br, and I): Novel optoelectronic materials showing bright emission with wide color gamut. *Nano Lett.* **15**, 3692-3696 (2015).

**Acknowledgements**

This work was supported by ETH Research Grant ETH-38 14-1, by the Swiss National Science Foundation under Grant Nr. 200021-140617, and by the U.S. Office of Naval Research (ONR) through the Naval Research Laboratory's Basic Research Program (SCE). F.D.O. benefitted from an ONR Global travel grant. S.J.P.K. acknowledges funding from the European Research Council under the European Union's Seventh Framework Programme (FP/2007-2013) / ERC Grant Agreement Nr. 339905 (QuaDoPS Advanced Grant). Computations were performed at the ETH High-Performance Computing Cluster Euler and the DoD Major Shared Resource Center at AFRL. We thank A. Sánchez-Ferrer for assistance with the WAXS measurements and R. Mezzenga for equipment access. We acknowledge L. Frenette, O. Hirsch, P. Kumar, D. Koziej, V. Lin, M. Mazzotti, K. McNeill, S. Meyer, M. Niederberger, F. Rabouw, A. Rossinelli, H. Schönberg, and O. Waser for technical assistance and discussions. We utilized the ScopeM facility at ETH Zurich for electron microscopy.

**Author contributions**

A.R. and F.D.O. contributed equally. A.R., F.D.O., S.C.E., and D.J.N. conceived the experiments and model. Syntheses and optical spectroscopy were performed by A.R., A.M., and P.N.K. X-ray diffraction, TGA/DSC-MS, nuclear magnetic resonance spectroscopy, electron microscopy, and energy-dispersive X-ray spectroscopy were done by A.R. Polarized optical microscopy was performed by A.R. with help from S.J.P.K and F.P. Calculations and simulations were performed by F.D.O and S.M. The NPL growth model was developed by F.D.O and S.C.E with input from A.R. and D.J.N. Both S.C.E.

and D.J.N. supervised the work. A.R., F.D.O., S.C.E. and D.J.N. wrote the manuscript. All authors contributed to the discussion of the results and to the revision of the manuscript.

**Additional information**

Supplementary information is available in the online version of the paper. Reprints and permissions information is available online at www.nature.com/reprints. Correspondence and requests for materials should be addressed to S.C.E. or D.J.N.

**Competing financial interests**

The authors declare no competing financial interests.

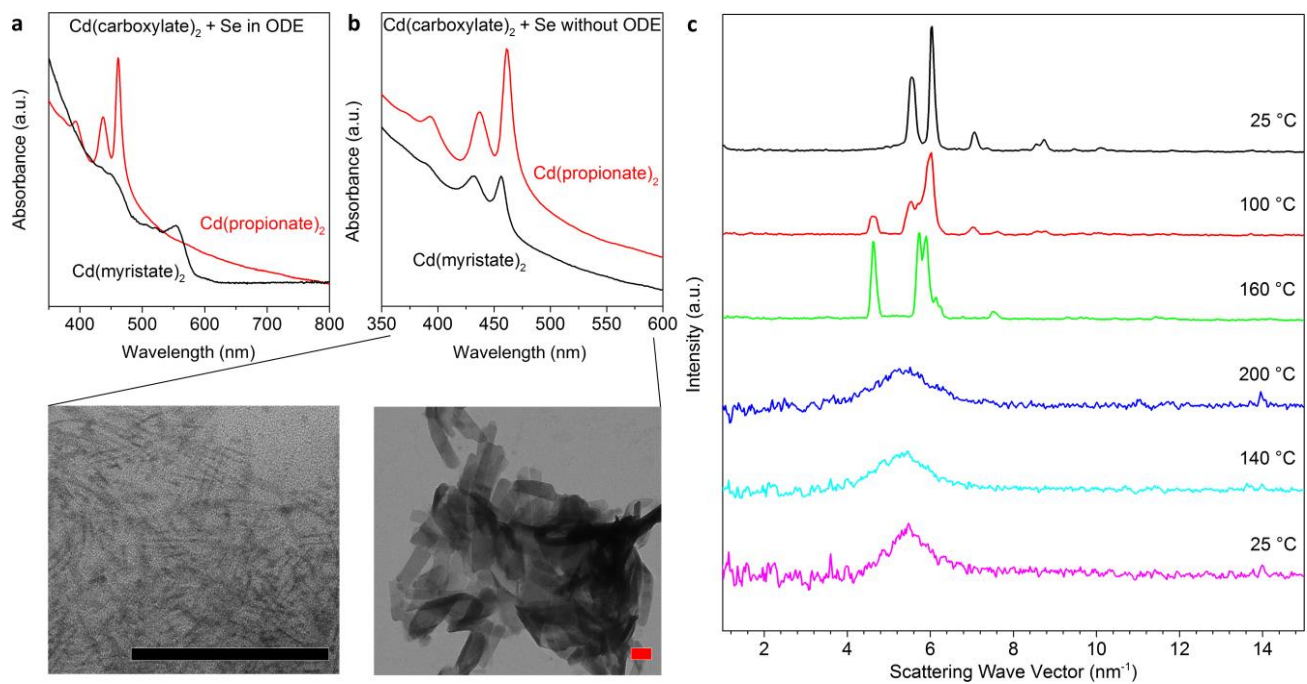

**Figure 1 | The formation of CdSe nanoplatelets in isotropic melts of Cd(carboxylate)$_2$ and elemental Se. a**, The solubility of the Cd(carboxylate)$_2$ in octadecene (ODE) determines whether quantum dots or nanoplatelets (NPLs) are formed, as shown via optical absorption. While Cd(myristate)$_2$ dissolved in ODE and yielded quantum dots with the lowest energy peak at 554 nm (black curve), Cd(propionate)$_2$ was insoluble above its melting point, and 4-monolayer CdSe NPLs were obtained (red curve), as indicated by the peaks at 437 and 461 nm. **b**, Without solvent, 4-monolayer CdSe NPLs formed in melts of either Cd precursor (over 18 h at 200 °C). Absorption peaks occurred at 432 and 457 nm for Cd(myristate)$_2$ and 436 and 462 nm for Cd(propionate)$_2$. Transmission electron micrographs show small NPLs obtained from Cd(myristate)$_2$ (left) and much larger NPLs from Cd(propionate)$_2$ (right). Both scale bars are 100 nm. **c**, *In situ* wide-angle X-ray scattering experiments reveal that the initially polycrystalline Cd(propionate)$_2$ undergoes a phase transition around 100 °C, becomes completely isotropic above the melting point, and maintains this glassy phase upon cooling.

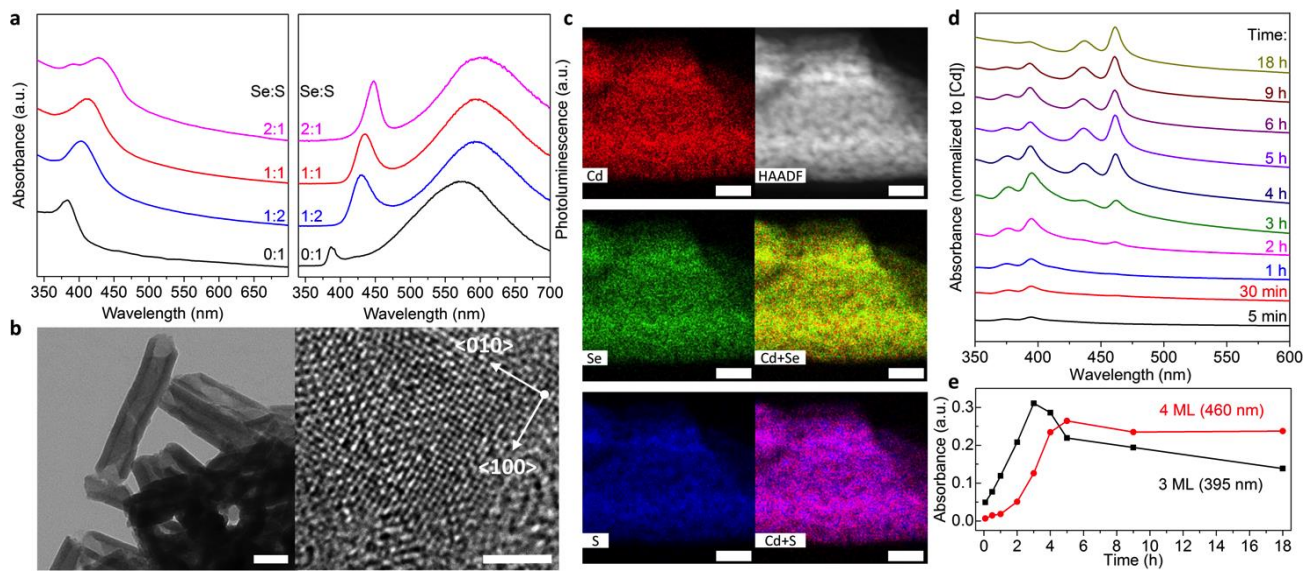

**Figure 2 | CdS$_{1-x}$Se$_x$ nanoplatelets and thickness transitions in CdSe nanoplatelets. a**, Optical absorption (left) and photoluminescence (right) spectra of CdS$_{1-x}$Se$_x$ nanoplatelets (NPLs) obtained with different Se to S ratios. The lowest energy absorption band (382 nm for pure CdS NPLs[11]) is shifted to lower energies with increasing Se content. Absorption and emission lines are broadened compared to pure CdSe or CdS NPLs. The large Stokes shift indicates alloying of Se and S. **b**, Transmission electron micrographs of CdS$_{1-x}$Se$_x$ NPLs obtained with equimolar amounts of Se and S. The large sheets role up presumably due to lattice strain. Scale bars are 50 and 2 nm. **c**, Energy dispersive X-ray analyses of the NPLs from **b** reveal a homogeneous distribution of Se and S. A high-angle annular dark field (HAADF) image is also shown for comparison. Scale bars are 10 nm. **d**, Optical absorption spectra taken over 18 h from a blend of Cd(propionate)$_2$ and Se powders at 200 °C show that at first 3-monolayer CdSe NPLs form, then slowly disappear with the appearance of 4-monolayer NPLs. **e**, The peak absorbance of the lowest energy feature from 3- and 4-monolayer NPLs from **d** plotted versus time.

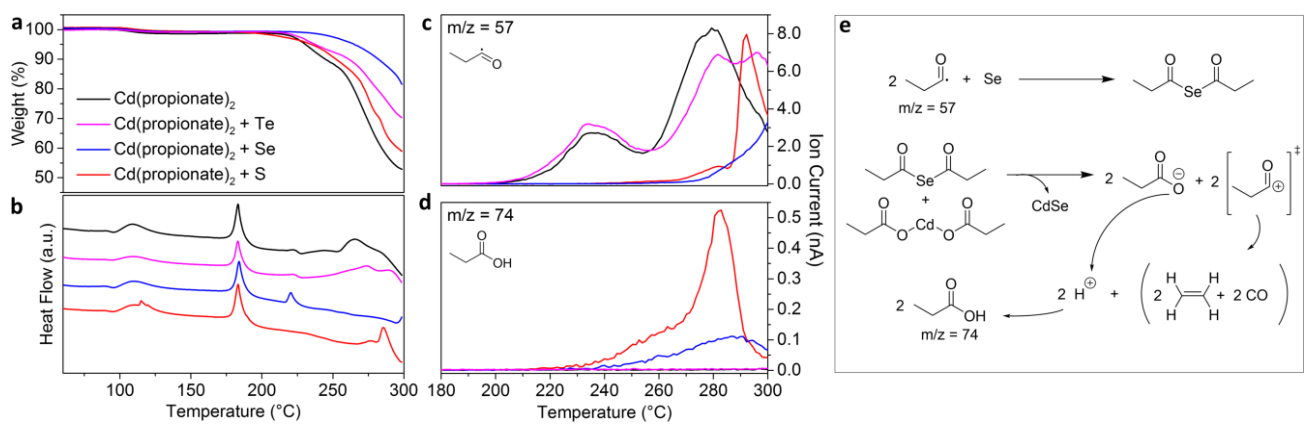

**Figure 3 | *In situ* analysis of the reactive chalcogenide species formed in melts of Cd(propionate)$_2$.
a**, Thermogravimetric analysis (TGA) and **b**, differential scanning calorimetric (DSC) data for Cd(propionate)$_2$ alone or with one equivalent of S, Se, or Te. Cd(propionate)$_2$ alone or with Te showed exothermic decomposition at 225 °C, absent for Cd(propionate)$_2$ with Se or S, which instead showed only melting of Se (220 °C) or S (115 °C). The endothermic melting of Cd(propionate)$_2$ occurred in all samples (180 °C). **c,d**, During TGA/DSC measurements, volatile species were analyzed by electron-ionization mass spectroscopy (MS). Below 250 °C, propionyl radicals (M$^+$, mass-to-charge, m/z, of 57) were detected only for Cd(propionate)$_2$ alone or with Te. Release of propionic acid (M$^+$, m/z=74) was observed only with Se or S. **e**, The reaction mechanism consistent with TGA/DSC/MS data. Decomposition of Cd(propionate)$_2$ yields propionyl radicals, which are captured by Se (or S). The resulting dipropionyl selenide (sulfide) reacts with Cd(propionate)$_2$ to produce CdSe (CdS) NPLs, releasing propionate anions and propionyl cations. The latter decompose to ethane, carbon monoxide, and protons. Free propionate can capture these protons. Indeed, propionic acid was detected only in samples containing Se (or S).

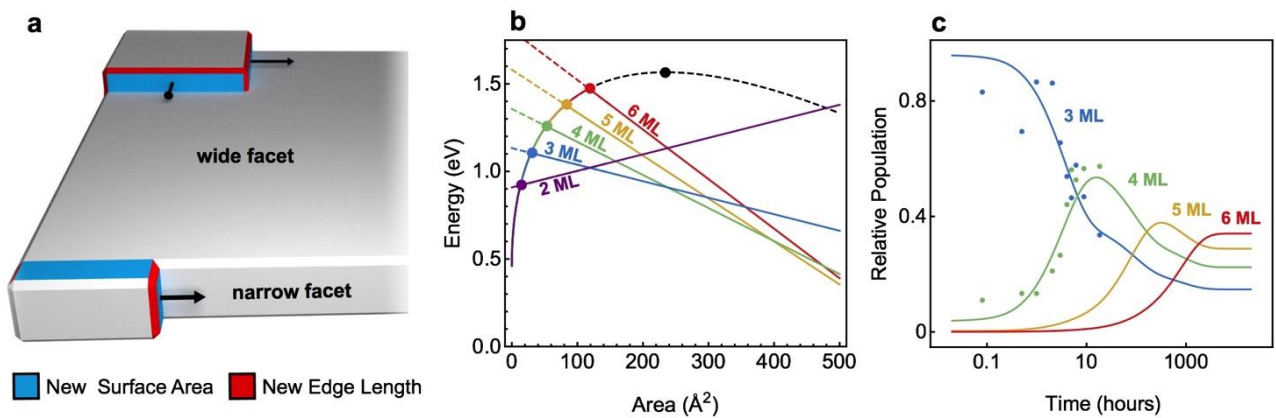

**Figure 4 | Theory of the intrinsic growth instability leading to nanoplatelets. a**, Qualitatively different growth modes on wide and narrow facets of a nanocrystal. On large facets a nucleated island grows isotropically to minimize its energy. The corresponding nucleation barrier is determined by the critical island size. On narrow facets with thickness less than this critical size, the island quickly spans the entire facet and then grows along the facet. The corresponding nucleation barrier is substantially reduced, leading to much faster growth at the experimental temperatures. Blue and red areas show new surfaces and edges formed by the islands, respectively. **b**, Calculated energy vs. island size on facets of different thicknesses (black: wide facets, color: narrow facets 2 to 6 monolayers) based on parameters obtained from DFT calculations. The corresponding nucleation barriers are marked by circles. **c**, Time evolution of the relative populations of nanoplatelets with thicknesses from 3 to 6 monolayers, obtained from reversible first-order kinetics and Arrhenius reaction rates evaluated from the nucleation barriers in **b**. A single attempt frequency, which sets the overall time scale, was fit to experimental populations of nanoplatelets 3- (blue) and 4-monolayers (green) thick.

# How semiconductor nanoplatelets form

Andreas Riedinger[1†], Florian D. Ott[1†], Aniket Mule[1], Sergio Mazzotti[1], Philippe N. Knüsel[1], Stephan J. P. Kress[1], Ferry Prins[1], Steven C. Erwin[2]* and David J. Norris[1]*

[1]Optical Materials Engineering Laboratory, Department of Mechanical and Process Engineering, ETH Zurich, 8092 Zurich, Switzerland.
[2]Center for Computational Materials Science, Naval Research Laboratory, Washington, D.C. 20375 USA.

[†]These authors contributed equally to this work.
*email: steve.erwin@nrl.navy.mil; dnorris@ethz.ch

## Supplementary information

### S1. Materials

**a. Chemicals.** Cadmium propionate (anhydrous) was purchased from MP Biomedicals (99.9%, #05209512). Cadmium nitrate tetrahydrate (99.9%, #20911), cadmium acetate dihydrate (98%, #289159), mercury acetate (98%, #176109), selenium (99.5%, #209651), sulfur (99.5%, #84683), myristic acid (98%, #70082), butyric acid (99%, #B103500), propionic acid (99.5%, #402907), acetic anhydride (99%, #320102), sodium hydroxide (98.5%, #06306), 1-octadecene (90%, #O806), oleic acid (90%, #364525), and chloroform-D (99.8 atom% deuterium, #151823) were purchased from Sigma Aldrich. Hexane and methanol were purchased from Thommen-Furler AG, and absolute ethanol from Alcosuisse AG. 1-Octadecene was degassed under vacuum (0.02 mbar) for 20 h at 100 °C and stored under nitrogen. All other chemicals were used as received.

### S2. Characterization

**a. Nuclear magnetic resonance spectroscopy (NMR).** Samples were dissolved in deuterated chloroform and transferred to an NMR tube. All spectra were recorded on a Bruker Ascend Aeon 400 MHz spectrometer. $^1$H-NMR spectra were recorded with predefined pulse programs. DOSY experiments were carried out with a first gradient amplitude of 2% and a final gradient amplitude of 95% with a linear ramp (32 points).

**b. Powder X-ray diffraction (XRD).** Samples were dispersed in hexane and drop-casted on zero background Si sample holders, letting the solvent evaporate. XRD patterns were recorded on a Bruker D8 Advance instrument (40 kV, 40 mA, $\lambda_{CuK\alpha}$ = 0.15418 nm).

**c. Wide angle X-ray scattering experiments (WAXS)** experiments were performed using a Rigaku MicroMax-002+ microfocused beam (40 W, 45 kV, 0.88 mA) with the $\lambda_{CuK\alpha}$ = 0.15418 nm radiation in order to obtain direct information on the scattering patterns. The scattering intensities were collected by a Fujifilm BAS-MS 2025 imaging plate system (15.2 cm × 15.2 cm, 50 µm resolution). An effective scattering-vector range of 1 nm$^{-1}$ < q < 15 nm$^{-1}$ was obtained, where q is the scattering wave vector defined as q = $4\pi \sin \theta / \lambda_{CuK\alpha}$ with a scattering angle of $2\theta$. Samples were placed in a Linkam THMS600 hot stage holder where temperature was set by a Linkam TMS94 temperature controller.

**d. Polarized optical microscopy (POM).** Materials were studied with an optical Zeiss Axiophot microscope equipped with crossed polarizers connected to a Linkam LTS 350 hot stage. Photomicrographs were taken with a Pixelink PL-A662 CMOS camera. Samples were prepared in a nitrogen-filled glovebox by placing a few micrograms of thoroughly mixed Cd(propionate)$_2$ and Se powders (molar ratio 3:1) between two standard microscope coverslips which were sealed with UV-curable epoxy to avoid air exposure during the experiment. The samples were removed from the glovebox, placed on the heating stage on the microscope, and studied in a temperature range from 25 to 200 °C with a heating rate of dT/dt = 10 K/min.

**e. Differential scanning calorimetry (DSC)** experiments were performed on a Mettler-Toledo DSC1 calorimeter equipped with a Huber TC100 cooling system. Samples (approximately 10 mg) were encapsulated in 40 µL aluminum crucibles and measured under nitrogen atmosphere with a heating/cooling rate of dT/dt = 10 K/min.

**f. Thermogravimetric analysis (TGA)/DSC coupled with mass spectrometry (MS).** Cd(propionate)$_2$ alone or with one molar equivalent of Te, Se, or S was weighed in Al$_2$O$_3$ crucibles and loaded in a Netzsch STA 449 thermal analyzer. After keeping the samples under an argon flux for 30 min to remove residual air, the temperature was ramped up from 45 to 300 °C at 10 K/min under argon flow (50 mL/min). TGA and DSC profiles were recorded in synchronization with electron ionization MS of volatile decomposition products.

**g. Absorption and photoluminescence spectroscopy.** Aliquots were diluted in hexane and transferred to quartz cuvettes. Absorption spectra were recorded on a Varian Cary 50 spectrophotometer and photoluminescence spectra on a Spex Fluorolog 2 fluorometer.

**h. Transmission electron microscopy (TEM) and energy-dispersive X-ray spectroscopy (EDS).** Samples were prepared on carbon-coated copper grids by drop-casting dilute dispersions (hexane) followed by slow evaporation of the solvent. TEM micrographs were recorded on a Philips CM12 (operated at 100 kV) or on a FEI Tecnai F30 FEG (operated at 300 kV). EDS analyses was carried out on a FEI Talos F200X microscope (operated at 200 kV).

**i. Scanning electron microscopy (SEM) and energy-dispersive X-ray spectroscopy (EDS).** Dispersions of powders (hexane) were drop-casted onto Si wafers and imaged using a Hitachi S-4800 SEM, operated at 3 kV in secondary-electron mode. EDS spectra were recorded at 10 kV with an EDAX Octane Super detector.

**S3.    Synthesis of materials**

**a. Cd(myristate)$_2$ and Cd(oleate)$_2$.** 3.54 g (15 mmol) Cd(NO$_3$)$_2$ · 4 H$_2$O and 6.85 g (30 mmol) myristic acid [or 8.47 g (30 mmol) oleic acid] were dissolved in 650 mL methanol under magnetic stirring. A solution of 1.81 g (45 mmol) NaOH in 50 mL methanol was added dropwise and left under vigorous stirring for 30 min. Cd(myristate)$_2$ [or Cd(oleate)$_2$] precipitated as a white, fluffy solid. This product was filtered, washed three times with methanol, and dried overnight in a vacuum oven at 40 °C (10 mbar).

**b. CdSe nanoplatelet (NPL) syntheses with acetic anhydride, propionic, or butyric acid.** The general protocol from Ref. S1 was adapted and modified. Briefly, in a 250 mL three-neck flask, 90 mL of 1-octadecene (ODE), 480 mg (1.8 mmol) Cd(acetate)$_2$ · 2 H$_2$O, and 1.18 g (4.2 mmol) technical grade oleic acid were degassed at 110 °C for 90 min under vacuum (0.02 mbar) and then kept under a nitrogen atmosphere. A dispersion of 72 mg (0.9 mmol) Se in 2 mL degassed ODE was prepared in a nitrogen-filled glove box and injected into the reaction vessel. The temperature was set to 240 °C with a ramp of 20 °C/min. When the reaction mixture reached 180 °C, 0.5, 1, or 2 equivalents (with respect to Cd) of either acetic anhydride, propionic acid, or butyric acid were injected. Aliquots (1 mL) were taken at 175 (before injection), 190, 205, 220, and 240 °C and diluted in 5 mL hexane. The temperature was maintained for 10 min at 240 °C, another aliquot was taken, and the mixture was quickly cooled to room temperature with a water bath and 10 mL oleic acid was injected. All products were precipitated from the crude reaction mixture with 5:1 excess of ethanol and centrifugation at 4000 r.p.m. (15 min). Special care was taken to ensure that the supernatants did not contain any CdSe species (quantum dots or NPLs) in order to get a representative overview of the products. The pellets were redispersed in hexane and precipitated two more times with ethanol by centrifugation (4000 r.p.m., 5 min). The washed products were redispersed and stored in 10 mL hexane.

**c. Studies on ligand exchange in Cd(oleate)$_2$ complexes.** A modified version of the protocol described in Section S3.b was used, *i.e.* without Se. In a 250 mL three-neck flask, 90 mL ODE, 480 mg (1.8 mmol) Cd(acetate)$_2$ · 2 H$_2$O, and 1.18 g (4.2 mmol) technical grade oleic acid were degassed at 110 °C for 90 min under vacuum (0.02 mbar) and then kept under a nitrogen atmosphere. At this stage, all Cd(acetate)$_2$ was converted to Cd(oleate)$_2$. The temperature was set to 180 °C with a ramp of 20 °C/min. When the solution reached 180 °C, 2 equivalents (with respect to Cd) of either acetic anhydride, propionic acid, or butyric acid were injected. Upon injection of acetic anhydride or propionic acid the solution became turbid, while in the case of butyric acid no changes were observed. The reaction was allowed to continue for 30 min at 180 °C, then the mixture was cooled to room temperature. Solids were isolated by direct centrifugation of the reaction mixtures at 8000 r.p.m. (30 min), washed three times with hexane, isolated by centrifugation (8000 r.p.m., 10 min), and dried under vacuum (0.02 mbar) for 24 h.

**d. Solvent-free synthesis of CdSe and CdS$_{1-x}$Se$_x$ NPLs.** Powders of anhydrous cadmium carboxylates [Cd(propionate)$_2$ or Cd(myristate)$_2$] were mixed with powders of Se, S, or mixtures of Se and S in glass vials at a molar ratio of 3:1 (cadmium carboxylates to chalcogens, if not stated otherwise) and placed in a pre-heated muffle oven inside a nitrogen-filled glovebox. Reactions were carried out from 5 min to 18 h at 200 °C (if not stated otherwise). The initially black (yellow) melt of cadmium carboxylates and Se (S) turned yellow (white) over time. After cooling to room temperature, the samples were taken from the glovebox, and solutions of oleic acid (5 vol%) in hexane/ethanol (1:1 vol%) were added to the dry powders, followed by sonication (30 min) and centrifugation at 3000 r.p.m. (5 min). The supernatants were discarded and the pellets were resuspended in ethanol, sonicated (30 min), and then centrifuged at 4000 r.p.m. (5 min). This redispersion-precipitation-centrifugation sequence was repeated twice, and the cleaned samples were stored in hexane.

**e. Solvent-free synthesis of mercury selenide.** 300 mg Hg(acetate)$_2$ (0.94 mmol) was mixed with 25 mg Se (0.32 mmol) in a glass vial and heated to 190 °C for 16 h inside a nitrogen-filled glovebox. The black reaction cake was dispersed in a solution of 5 mL hexane and 200 μL oleic acid followed by sonication (30 min). The sample was precipitated by centrifugation at 4000 r.p.m. (5 min), redispersed in hexane, precipitated with an excess of ethanol, and centrifuged again at 4000 r.p.m. (5 min). This redispersion-precipitation-centrifugation sequence was repeated twice, and the washed sample was dispersed in hexane.

## S4. Modeling and theory

**a. Model of two-dimensional nucleation and growth of nanoplatelets.** Here we provide further details about our evaluation of the nucleation barriers to form islands on surfaces. We assume that the change in energy upon island nucleation depends on the change in total crystal volume ($\Delta V$), surface area ($\Delta A$), and edge length ($\Delta L$):

$$\Delta E = E_V \Delta V + E_A \Delta A + E_L \Delta L. \tag{S1}$$

Here, $E_V$, $E_A$, and $E_L$ are respectively volume, surface, and line energies. $E_V$ is the energy per unit volume of CdSe in the zincblende bulk crystal phase, relative to its value in the precursor melt; thus $E_V$ is the thermodynamic driving force for crystallization. $E_A$ is the surface formation energy of CdSe(001) passivated by Cd-acetate molecules. From DFT calculations (see Section S4.c) we obtained $E_A$= 5.7 meV/Å² under Se-rich conditions. $E_L$ is the edge formation energy of a monolayer-high step on Cd(001); from DFT we obtained 37.1 meV/Å. We consider only monolayer islands, for which the height is $L_1 = \frac{a_0}{2} = 3.10$ Å. Hence the additional volume $\Delta V$ for an island is equal to the projected area, $a$, of the island times the island height: $\Delta V = L_1 a$. Similarly, the additional surface area $\Delta A$ is equal to the projected perimeter of the island ($L_\text{island}$) times the island height: $\Delta A = L_1 L_\text{island}$. The evaluation of the total added edge length is slightly more complex. As can be seen from Fig. 4a in the main text, the increase in total edge length arises from the part of the island perimeter forming a step with length $L_\text{step}$, plus 4 edges with length $L_1$: $\Delta L = L_\text{step} + 4L_1$. We substitute these expressions into Eq. (S1) and obtain for the island formation energy:

$$\Delta E(a, L_\text{island}, L_\text{step}) = (L_1 E_V)a + L_1 E_A L_\text{island} + E_L(L_\text{step} + 4L_1). \tag{S2}$$

The shape of a growing island evolves so as to minimize its formation energy at every point along a generalized reaction coordinate which we take to be the projected area of the island:

$$\text{d}\Delta E|_a = 0. \tag{S3}$$

The solution of Eq. (S3) describes the minimum energy path for the formation reaction of a stable island. The maximum of this curve, if it exists, corresponds to the nucleation barrier, $\Delta E^{\text{barrier}}$. We evaluate the formation energy of the most stable island with a given area and underlying surface shape under our assumption that the crystallite is bounded by {001} facets. We find that the solution of Eq. (S3) depends on the facet dimensions. On wide facets, for which the characteristic size of the island is smaller than the facet thickness, we assume for simplicity a square island originating from a corner of the facet (see Fig. 4a in the main text). The formation energy is

$$E^{\text{wide}}(a) = L_1 E_V a + 4L_1 \sqrt{a} E_A + 2(\sqrt{a} + 2L_1)E_L \tag{S4}$$

$$= L_1 E_V a + (4L_1 E_A + 2E_L)\sqrt{a} + 4L_1 E_L. \tag{S5}$$

On narrow facets, the most stable island has a rectangular shape covering the full width of the facet (see Fig. 4a in the main text). In this regime the island formation energy is

$$E^{\text{narrow}}(a) = L_1 E_V a + \left(2mL_1 + \frac{2a}{mL_1}\right) L_1 E_A + (m+4)L_1 E_L. \tag{S6}$$

$$= \left(L_1 E_V + \frac{2E_A}{m}\right) a + 2mL_1^2 E_A + (m+4)L_1 E_L. \tag{S7}$$

These solutions are shown in Fig. 4b in the main text for a range of different surface thicknesses *m*.

**b. Simple kinetic model of reversible crystal growth.** The above nucleation model explains why growth is faster on thin facets, assuming an overall driving force for growth exists. However, it does not take into account the reversibility of the growth or the fact that thermodynamic driving forces may change throughout the reaction, for example, by the decaying supersaturation of reactants. To address this, we developed a simple kinetic model to describe the temporal evolution of five different populations: free monomers ($N_0$), and monomers embedded in NPLs with thicknesses from 3 to 6 monolayers ($N_{3-6}$). NPLs thinner than 3 monolayers are not stable, and thicker ones grow with rates far beyond laboratory timescales.

To begin, we constrain our system by the mass conservation law:

$$\sum_m N_m(t) = N_{\text{tot}} . \tag{S8}$$

We describe attachment and detachment reactions of monomers on the crystal surfaces as first-order processes. This means that we assume that on a nanoplatelet ($m = 3,4,5,6$) the attachment rate is proportional to the free-monomer concentration and the detachment rate is proportional to the bound-monomer concentration:

$$\frac{dN_m(t)}{dt} = k_m N_0(t) - k_{-m} N_m(t) , \tag{S9}$$

where $k_m$ and $k_{-m}$ are the rate constants for attachment and detachment, respectively. Based on (S8) and (S9) we can also express $N_0(t)$ in its derivative form:

$$\frac{dN_0(t)}{dt} = -N_0(t) \sum_{m \neq 0} k_m + \sum_{m \neq 0} k_{-m} N_m(t) . \tag{S10}$$

We can express the set of differential equations in matrix notation ($\sum_{m \neq 0} k_m = k_{\text{tot}}$):

$$\frac{d}{dt} \begin{pmatrix} N_0(t) \\ N_3(t) \\ N_4(t) \\ N_5(t) \\ N_6(t) \end{pmatrix} = \begin{pmatrix} -k_{\text{tot}} & k_{-3} & k_{-4} & k_{-5} & k_{-6} \\ k_3 & -k_{-3} & 0 & 0 & 0 \\ k_4 & 0 & -k_{-4} & 0 & 0 \\ k_5 & 0 & 0 & -k_{-5} & 0 \\ k_6 & 0 & 0 & 0 & -k_{-6} \end{pmatrix} \begin{pmatrix} N_0(t) \\ N_3(t) \\ N_4(t) \\ N_5(t) \\ N_6(t) \end{pmatrix} , \tag{S11}$$

or written in compact form:

$$\frac{d}{dt} x(t) = \mathbf{K} x(t) . \tag{S12}$$

This problem can be diagonalized using the unitary matrix $\mathbf{U}$ which contains column-wise the eigenvectors of $\mathbf{K}$:

$$\frac{d}{dt}[\mathbf{U}^{-1} x(t)] = (\mathbf{U}^{-1} \mathbf{K} \mathbf{U})[\mathbf{U}^{-1} x(t)] , \tag{S13}$$

$$\frac{d}{dt}\widetilde{\boldsymbol{x}}(t) = \boldsymbol{\lambda}\widetilde{\boldsymbol{x}}(t) , \tag{S14}$$

where $\widetilde{\boldsymbol{x}}(t) = \mathbf{U}^{-1}\boldsymbol{x}(t)$ and $\boldsymbol{\lambda} = \mathbf{U}^{-1}\mathbf{K}\mathbf{U}$ is a diagonal matrix containing the eigenvalues of $\mathbf{K}$. Having uncoupled the set of differential equations they can be solved separately:

$$\frac{d}{dt}\widetilde{x}_i(t) = \lambda_{ii}\widetilde{x}_i(t) , \tag{S15}$$

$$\widetilde{x}_i(t) = \widetilde{x}_i(0)e^{\lambda_{ii}t} . \tag{S16}$$

To find the values $\widetilde{x}_i(0)$ we need to evaluate the coordinate transformation at $t = 0$ for where we know that all monomers ($N_{\text{tot}}$) are in their free form:

$$\widetilde{\boldsymbol{x}}(0) = \mathbf{U}^{-1}\boldsymbol{x}(0) = \mathbf{U}^{-1}(N_{\text{tot}}, 0,0,0,0)^T . \tag{S17}$$

Finally, we obtain the solution by reversing the transformation from above:

$$\boldsymbol{x}(t) = \mathbf{U}\widetilde{\boldsymbol{x}}(t) . \tag{S18}$$

We next need to evaluate $k_m$ and $k_{-m}$. Monomer attachment and detachment occur on kink sites along surface steps at the edges of islands[S2,S3,S4]. The number of such islands is proportional to the rate at which they are nucleated, $\exp\left(-\frac{\Delta E_m^{\text{barrier}}}{kT}\right)$, where $\Delta E_m^{\text{barrier}}$ is the island nucleation barrier. The rates of monomer attachment and detachment are proportional to the island number and the facet thickness $m$:

$$k_m = C_0 m \exp\left(-\frac{\Delta E_m^{\text{barrier}}}{kT}\right) \tag{S19}$$

$$k_{-m} = C_0 m \exp\left(-\frac{\Delta E_m^{\text{barrier}}}{kT}\right) A_m. \tag{S20}$$

The prefactor $C_0$ includes all terms in the rate constant that are assumed to be equal for the various reactions, and $A_m$ accounts for the additional energy penalty to detach a monomer from its bound state in the nanoplatelet, making the detachment reaction slower. Since the ratio $k_m/k_{-m}$ is equal to the equilibrium constant we find for $A_m$:

$$A_m = \frac{k_{-m}}{k_m} = \exp\left(\frac{\Delta H_{f,m}}{kT}\right) = \exp\left(\frac{2L_1^3 E_V + \frac{4L_1^2}{m} E_A}{kT}\right), \quad (S21)$$

with $\Delta H_{f,m}$ being the reaction energy per monomer.

The remaining free parameter, $C_0$, was used to fit the solution of this set of ordinary differential equations to the experimentally recorded absorption spectra (Fig. 4c in the main text) of 3- and 4-monolayer-thick NPLs (the absorption data was normalized to the total amount of CdSe absorption at 350 nm). We obtained a general attempt frequency of $C_0 = 2.8 \times 10^{11}$ Hz. The resulting dynamics of the different populations are shown in Fig. S13a and Fig. 4c in the main text.

We now consider our assumption of first-order kinetics for monomer incorporation and dissolution. Derivations of explicit nucleation rates reveal second-order kinetics[S5] and the rate of dissolution is proportional to the amount of exposed monomers on the reactive surface rather than to the total amount of monomers in the NPLs. Since the reactive surface involves the few-monolayer-thick side facets, its area is proportional to $N_m^{1/2}(t)$. These considerations lead to kinetics very different from first order:

$$\frac{dN_m(t)}{dt} = k_m N_0^2(t) - k_{-m} N_m^{1/2}(t) \:. \quad (S22)$$

In Fig. S13, we show a comparison of numerical solutions for our first-order-kinetics model and the advanced model based on the same set of reaction constants, $k_m$ and $k_{-m}$. We can clearly observe the same qualitative behavior in both cases. In the more realistic model, the equilibrium conditions have changed due to the reaction order:

$$K_m = \exp\left(-\frac{2L_1^3 E_V + \frac{4L_1^2}{m} E_A}{kT}\right) = \frac{N_{m,eq}^{1/2}}{N_{0,eq}^2}, \quad (S23)$$

which leads to an equilibrium ratio between populations $i$ and $j$ that is squared compared to the first-order model:

$$\frac{N_{i,\text{eq}}}{N_{j,\text{eq}}} = \frac{K_i^2}{K_j^2}. \tag{S24}$$

The important point is that the simplified model predicts the same qualitative behavior as the exact model. It only slightly underestimates the selectivity of the synthesis.

**c. Evaluation of surface, edge, and volume energies.** We used density-functional theory (DFT) total-energy calculations to determine the atomistic geometry of passivated nanocrystal surfaces, steps, and edges, as well as the binding energies of CdSe monomers and the activation barriers for desorption of Cd-acetate ligands at those surfaces. For computational efficiency we omitted carboxylates with longer carbon chains as they can be expected to behave very similarly to acetate. Total energies and forces were calculated within the generalized-gradient approximation of Perdew, Burke, and Ernzerhof (PBE)[S6,S7] using projector-augmented-wave potentials, as implemented in VASP[S8,S9]. The plane-wave cut-off was 300 eV. We optimized all geometries until the forces were below 50 meV/Å.

The surface and edge energies, $E_A$ and $E_L$, on CdSe(001) are formation energies defined with respect to a reference system (here the reservoirs of different chemical species),[S10,S11,S12]

$$\Delta H_f(X) = E_{\text{tot}}^{\text{DFT}}(X) - \sum_i n_i \mu_i. \tag{S25}$$

We use the superscript "DFT" for values directly taken from a single DFT total-energy calculation. $X$ is the system including the defect (a surface or a step). The sum in Eq. (S25) accounts for the fact that each element $i$ contained $n_i$ times in $X$ has a chemical potential, $\mu_i$.

The NPLs, surfaces, and edges are comprised of Cd, Se, and the surfactant acetate molecule, (AcO, $H_3C_2O_2$), which we consider to be indivisible. Following standard practice we assumed thermodynamic equilibrium between CdSe and reservoirs of Cd and Se. Thus, we have:

S11

$$(\mu_{Cd} + \mu_{Se}) = E_{CdSe}^{DFT}. \tag{S26}$$

The analogous assumption for Cd(AcO)₂ gives:

$$\mu_{Cd} + 2\mu_{AcO} = E_{Cd(AcO)_2}^{DFT}. \tag{S27}$$

Due to the presence of elemental selenium in the reaction we also assumed, for concreteness, that the experiments were conducted under Se-rich conditions:

$$\mu_{Se} = E_{Se,bulk}^{DFT} \tag{S28}$$

These three conditions fix the values of the three chemical potentials. For the bulk phases we used the relaxed structures of zinc-blende CdSe ($E_{CdSe}^{DFT}$), γ-Se ($E_{Se,bulk}^{DFT}$), and a 2-dimensional coordination polymer of Cd acetate ($E_{Cd(OAc)_2}^{DFT}$, see Fig. S14).

To calculate the surface energy $E_A$ we used a 4-monolayer thick CdSe slab in vacuum with Cd-terminated {001} surfaces on both sides passivated with acetate molecules (see Fig. S15). We obtained $E_A = 5.7 \frac{meV}{Å^2}$, the value used in the main text.

To define the edge energy $E_L$ we used three islands of different width (Fig. S16b-d) and took the average as a single representative value appropriate for small islands. This average is $E_L = 37 \frac{meV}{Å}$, the value used in the main text. We caution that the high and low values (47 and 24 meV/Å) are quite different from this average and hence that the results should be understood as trends rather than precise values.

Finally, we turn to the volume energy $E_V$. Crystals form from supersaturated melts or solutions because a thermodynamic driving force exists for a monomer in the solution/melt to incorporate into the crystal. In our model the energy gain per monomer incorporated into a bulk crystal is given by $V_m E_V$, where $V_m$ is the volume of a monomer in bulk CdSe. While $E_A$ and $E_L$ can be estimated from DFT calculations for well-defined structures, this is less straightforward for $E_V$ because the melt is difficult to model. Thus, instead of DFT, we used experimental observations to constrain the range of possible values for $E_V$.

As discussed in the main text, stable nanoplatelets can only grow if their formation energy is negative:

$$\frac{dE^{\text{narrow-facet}}(a)}{da} = L_1 E_V + \frac{2}{m} E_A < 0. \tag{S29}$$

Because we can grow NPLs with a thickness of 3 monolayers, but not 2, we can conclude that

$$-\frac{2}{2L_1} E_A < E_V < -\frac{2}{3L_1} E_A. \tag{S30}$$

In our model as described in the main text, we used the average of the limits for this range, $E_V = -\frac{5}{6L_1} E_A = -1.5 \, \frac{\text{meV}}{\text{Å}^3}$.

**d. Calculation of the surfactant desorption barriers.** We evaluated the activation barriers for desorption of the surfactants from the NPL surface to better understand where CdSe monomers are most likely to adsorb during growth. We found that monomers preferentially adsorb near steps, which is consistent with the growth model described in the main text.

To evaluate the barrier for an acetate (or Cd acetate) to desorb from the CdSe surface we determined the minimum-energy reaction path using the nudged-elastic-band (NEB) method as implemented in VASP[S13]. We used 9 NEB images between the two endpoints of the reaction coordinates; this identifies the transition state with sufficient accuracy for our purposes. The specific reactions we investigated included: (i) acetate desorption from a flat surface, (ii) Cd acetate desorption from a flat surface, (iii) acetate desorption from a step, and (iv) Hg acetate desorption from a flat HgSe surface. Initial and final structures and the corresponding minimum energy paths of these reactions are shown in Fig. S17.

**e. Kinetic Monte Carlo simulations of nanoplatelet growth.** Atomistic simulations of growth provide an independent way to validate our explanation of the instability that leads to NPLs. We used the kinetic Monte Carlo method[S14,S15] to simulate nanocrystal growth via the incorporation of CdSe monomers. For simplicity we represented the underlying crystal lattice as simple cubic. We considered only adsorption and desorption of monomers, but not their diffusion (either on the surface of the nanocrystal or in solution). We treated monomer adsorption and desorption as thermally activated processes with Arrhenius-type reaction rates.

In the real system, the adsorption of a CdSe monomer requires three separate steps: (1) desorption of a surfactant molecule to expose a surface site, (2) adsorption of a monomer onto that site, and (3) re-adsorption of a surfactant molecule to maintain the passivated surface. Of these three steps only the first has an energy barrier. Therefore, the *effective* barrier for adding a monomer to a surface site $\Delta E_{\text{ads}}^{\text{barrier}}$ is given by the barrier for desorbing a surfactant molecule. We obtained these barriers from DFT/NEB calculations as discussed in the previous section. The results showed a strong preference of monomer adsorption on a step edge ($\Delta E_{\text{ads}}^{\text{barrier}} = 1.2$ eV) compared to a flat crystal surface ($\Delta E_{\text{ads}}^{\text{barrier}} = 2.1$ eV).

The desorption of a CdSe monomer also requires three steps: (1) desorption of the surfactant molecule, (2) desorption of a monomer molecule, and (3) re-adsorption of a surfactant. Two of these steps have a barrier, namely desorption of the surfactant and of the monomer. Compared to surfactants, monomers have a local environment which is much more variable – with 6 nearest-neighbour sites already in a simple cubic structure – leading to a large number of distinct scenarios. Instead of investigating each of these scenarios by DFT, we used a simplified model to estimate these site-dependent desorption barriers. First, we assume that monomer and surfactant desorb together as a complex. Furthermore, we assume that the transition state of this reaction is reached when monomer and surfactant have broken their bond to the surface, but no surfactant has yet adsorbed at the open site left behind. As a consequence (see the schematic potential-energy surface in Fig. S18), the barrier for monomer desorption can be easily estimated from the barrier for adsorption if the energy difference $\Delta E_{\text{des}}$ between the two endpoints of the reaction is also known. Thus we have:

$$\Delta E_{\text{des}}^{\text{barrier}} = \Delta E_{\text{ads}}^{\text{barrier}} + \Delta E_{\text{des}}. \tag{S31}$$

It is straightforward to evaluate the energy difference $\Delta E_{\text{des}}$ for any given local environment of a monomer. To do this, we used a modified bond-counting model based on a lattice of monomers. This model gives desorption barriers ranging from 0.075 eV for a single monomer sitting on a corner of a flat surface to 2.2 eV for a monomer incorporated in a flat surface. Each monomer in the nanocrystal makes $n_B$ bonds (from 1 to 6) to its nearest neighbors. When a monomer is removed from the nanocrystal these bonds are broken and the monomer is considered "dissolved" in the melt (or solvent). Hence, desorption raises the energy of the system by $n_B \varepsilon_B$ and lowers it by $|\mu|$, where $\mu < 0$ is the chemical potential of the dissolved monomer. We modified this standard bond-counting model by including an additional term arising from the energy per unit length $E_L$ of the exposed edges of the nanocrystal; this parameter has the same value as in the main text, $E_L = 37 \frac{\text{meV}}{\text{Å}}$. Hence the energy change upon monomer desorption is

$$\Delta E_{\text{des}} = n_B \varepsilon_B + \mu_S + E_L \Delta L. \tag{S32}$$

Finally, from expression (S31) follows a standard Arrhenius relationship between $r_{\text{des}}$ and $r_{\text{ads}}$[S2,S3,S4]:

$$r_{\text{des}} = r_{\text{ads}} \exp\left(-\frac{\Delta E_{\text{des}}}{kT}\right). \tag{S33}$$

We now show that by enforcing the equivalence of the discrete expression (S32) with our continuum model (in the main text) for the energy of a nanocrystal, we can obtain the parameters $\varepsilon_B$ and $\mu_S$ from the continuum parameters $E_V$ and $E_A$.

To do this, we equate the two different expressions for the energy of a general crystallite built from monomer cubes with the volume $V_m$. We begin with the continuum description. We group the monomers comprising the crystallite according to the number of their nearest neighbors: $n_1$ monomers with one nearest neighbor, $n_2$ monomers with two nearest neighbors, etc. Monomers with fewer than six nearest neighbors lead to a monomer face being exposed at the crystallite surface. Hence a monomer with $j$ nearest neighbors has $(6-j)$ exposed square surfaces, each with area $V_m^{2/3}$. Accordingly, the total energy is

$$E_{\text{continuum}} = \sum_{i=1}^{6} n_i \left[ V_m E_V + (6-i) V_m^{2/3} E_A \right] + E_L L, \tag{S34}$$

where $L$ is the total edge length of the crystallite.

We now make the same evaluation using the discrete expression (S32). Each monomer in the crystallite must first be obtained from the melt and hence costs energy $\mu_S$. In the crystallite, a monomer shares its bond with its $n_i$ nearest neighbors. Hence the total energy is

$$E_{\text{discrete}} = \sum_{i=1}^{6} n_i \left[ -\frac{i}{2} \varepsilon_B - \mu_S \right] + E_L L. \tag{S35}$$

In order for the expressions (S34) and (S35) to be equal for arbitrary choices of $\{n_i\}$ we must have

$$-\frac{i}{2} \varepsilon_B - \mu_S = V_m E_V + (6-i) V_m^{2/3} E_A, \tag{S36}$$

which immediately yields the parameters $\varepsilon_B$ and $\mu_S$ in terms of $E_V$ and $E_A$:

$$\varepsilon_B = 2 V_m^{2/3} E_S = 173 \text{ meV}, \tag{S37}$$

$$\mu_S = -V_m E_V - 6 V_m^{2/3} E_S = -428 \text{ meV}. \tag{S38}$$

We started our simulations from small random crystallite seeds of 6-28 monomers. At the actual experimental temperatures around 500 K the growth is extremely slow. This is because overgrowing a facet with a new layer starts with a sequence of highly unfavorable attachment steps. For example, the formation of a stable island on a 4-monolayer wide facet along the minimum energy path requires a sequence of 5 very rare attachment steps with probabilities on the order of $10^{-6}$, $10^{-2}$, $10^{-2}$, $10^{-1}$, and $10^{-4}$, respectively. Thus, to increase computational efficiency we ran the simulations at 1000 K, we blocked monomer attachment on flat surfaces due to its low reactivity (see surfactant desorption barriers), and we increased the sticking probability of single monomers to avoid repeated detachment and attachment steps. A typical growth sequence is shown in Movie S1 and a statistical summary of the results from small initial crystal seeds containing between 6 and 28 monomers is shown in Fig. S19. By increasing the temperature even further, for example to 2000 K, we observed that the selective growth of narrow facets is completely suppressed, and the crystallites grow isotropically (see Movie S2). These results are consistent with the 2D nucleation model presented in the main text.

Due to the high step energy of growing islands, the crystal facets in our simulations are not rough but almost perfectly flat: an island either quickly re-dissolves or overgrows the complete facet. Therefore, we can conveniently describe these crystals by the three side lengths of a rectangular box. We use the largest box spanned by the lattice points containing no unoccupied sites. We define the shortest side length as the thickness. Then the characteristic side length of the nanoplatelet is given by the square root of the maximum projected area of the box. In Fig. S19 we plot the thickness versus characteristic length for a set of crystals obtained from simulations starting with small crystal seeds of various shapes. We observe two types of growth: (1) crystals that reach an increased thickness at early stages in the simulations and grow equally slowly in all directions, forming isotropic crystallites; (2) crystals that remain thin while growing faster in the other two lateral dimensions, forming extended flat nanoplatelets.

Thus, these KMC simulations give insight into the microscopic mechanism of nanoplatelet growth. Using the simple concepts of nearest-neighbor binding and a penalty for forming edges, these atomistic simulations confirm our 2-dimensional nucleation and growth model. We anticipate that this method will allow us to further explore the parameter space controlling nanoplatelet growth and to investigate its behavior on extended time scales.

## S5. Supplementary figures

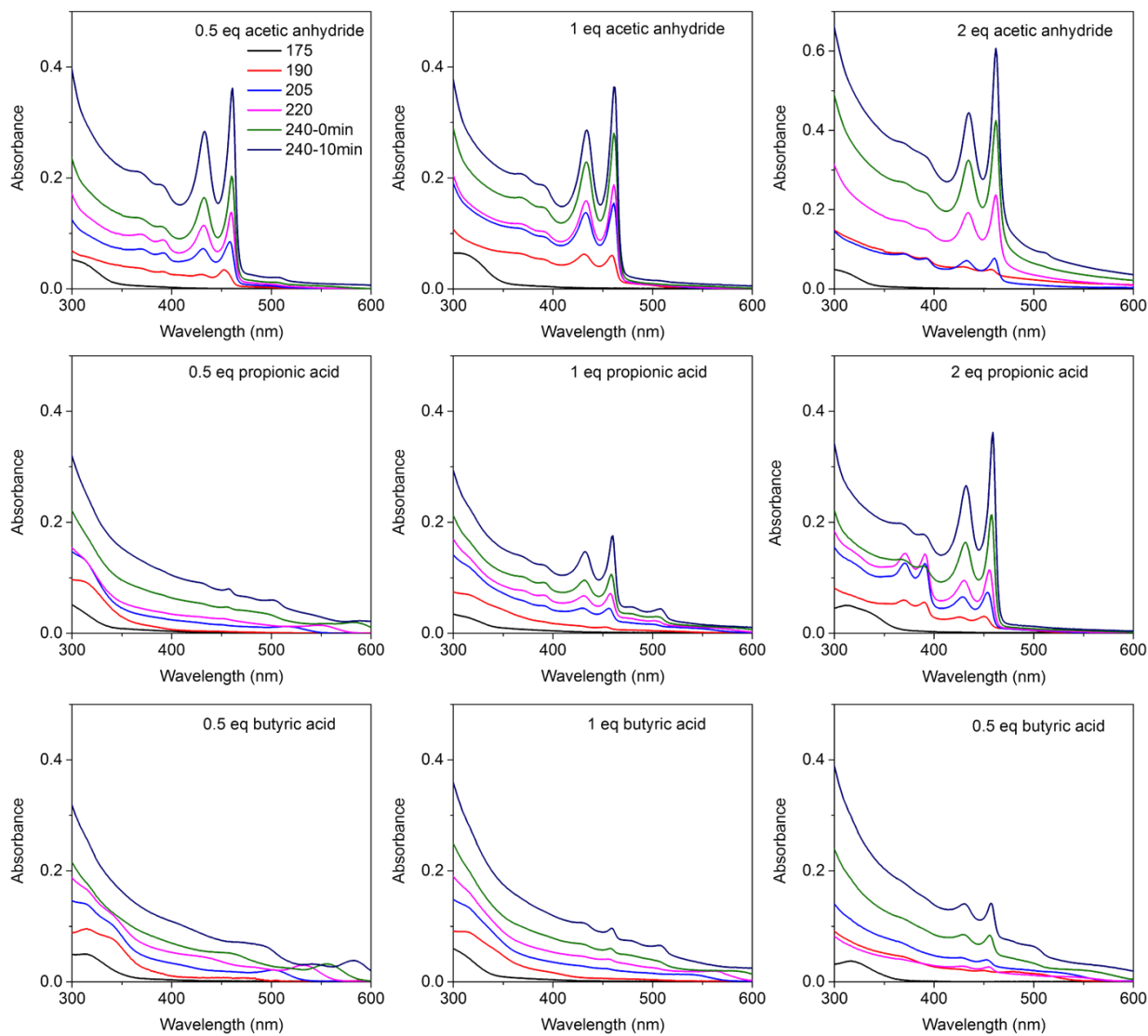

**Figure S1 | Absorption spectra taken during NPL syntheses with acetic anhydride, propionic acid, and butyric acid.** Spectra of the aliquots collected during the course of the reaction were recorded without further dilution or purification. The dilution factor was the same for all aliquots and syntheses, hence differences in absorbance between samples are absolute and thus directly comparable. A clear trend was observed: the longer the aliphatic chain of the carboxylic acids, the higher their required concentration for efficient NPL formation. The injection of acetic anhydride yielded NPLs even at 0.5 eq, propionic acid around 1 eq, and butyric acid yielded few NPLs even at 2 eqs. When 2 eq acetic anhydride were injected, significant scattering was observed. Interestingly, only with 2 eq propionic acid, clear absorption features of 3ML NPLs (371 nm and 391 nm) were observed at early time points, which diminished over the course of the reaction while absorption intensities of 4ML NPLs (432 nm and 458 nm) increased and were dominant at the end of the reaction.

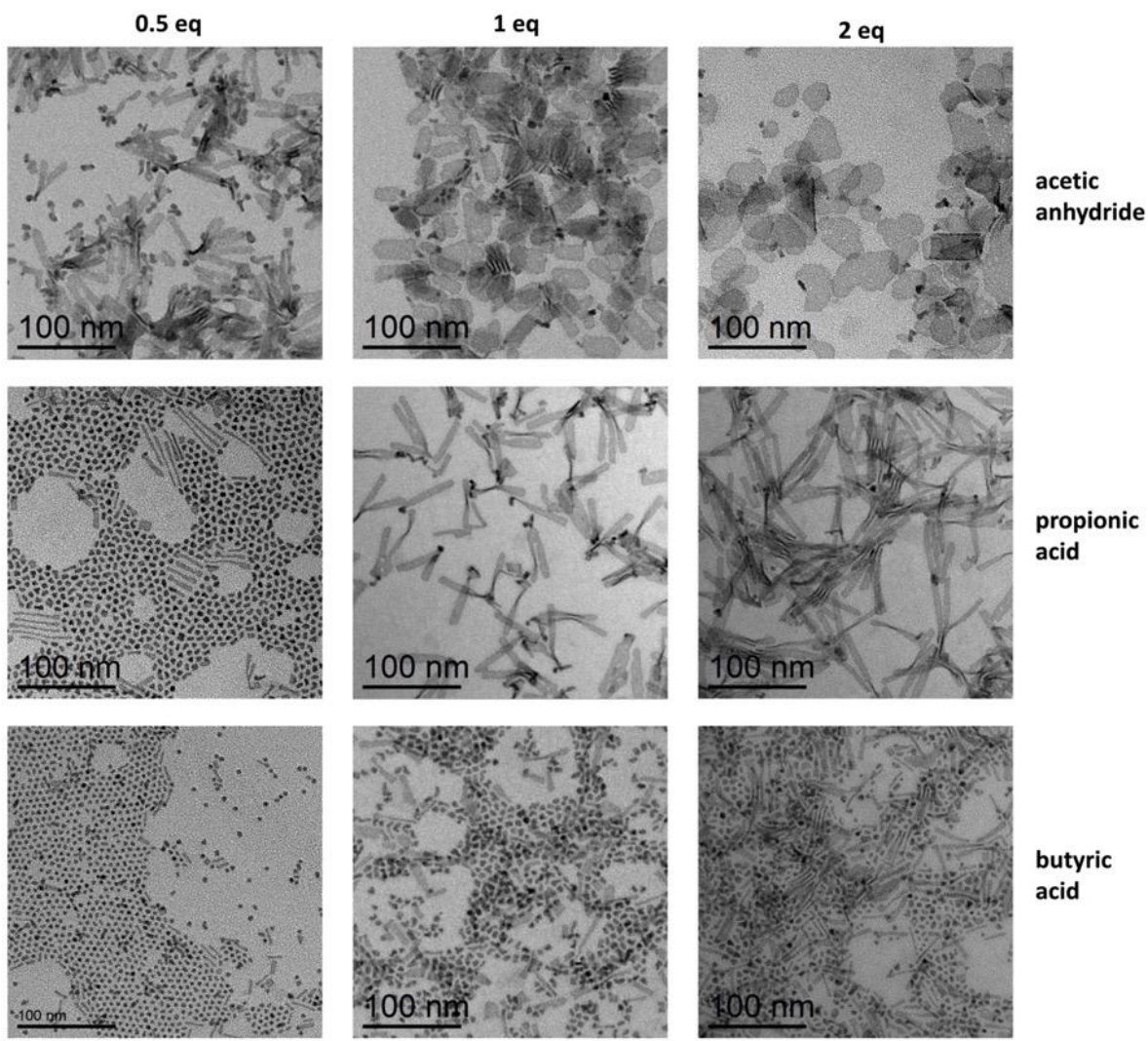

**Figure S2 | TEM micrographs of the reaction products obtained with 0.5 to 2 eq acetic anhydride, propionic acid, and butyric acid.** Lateral dimensions of NPLs increased with acetic anhydride concentration, while the aspect ratio decreases, and more irregularly shaped NPLs were formed with 2 eq. 0.5 eq propionic acid yielded mainly quasi-zero-dimensional nanocrystals (quantum dots) and just few NPLs with high aspect ratio were observed, which did not change much as the concentration of propionic acid was increased (even though lateral dimension increased). The addition of butyric acid yielded only a few NPLs at all tested concentrations. At the highest concentration of butyric acid, more irregularly shaped nanocrystals were observed.

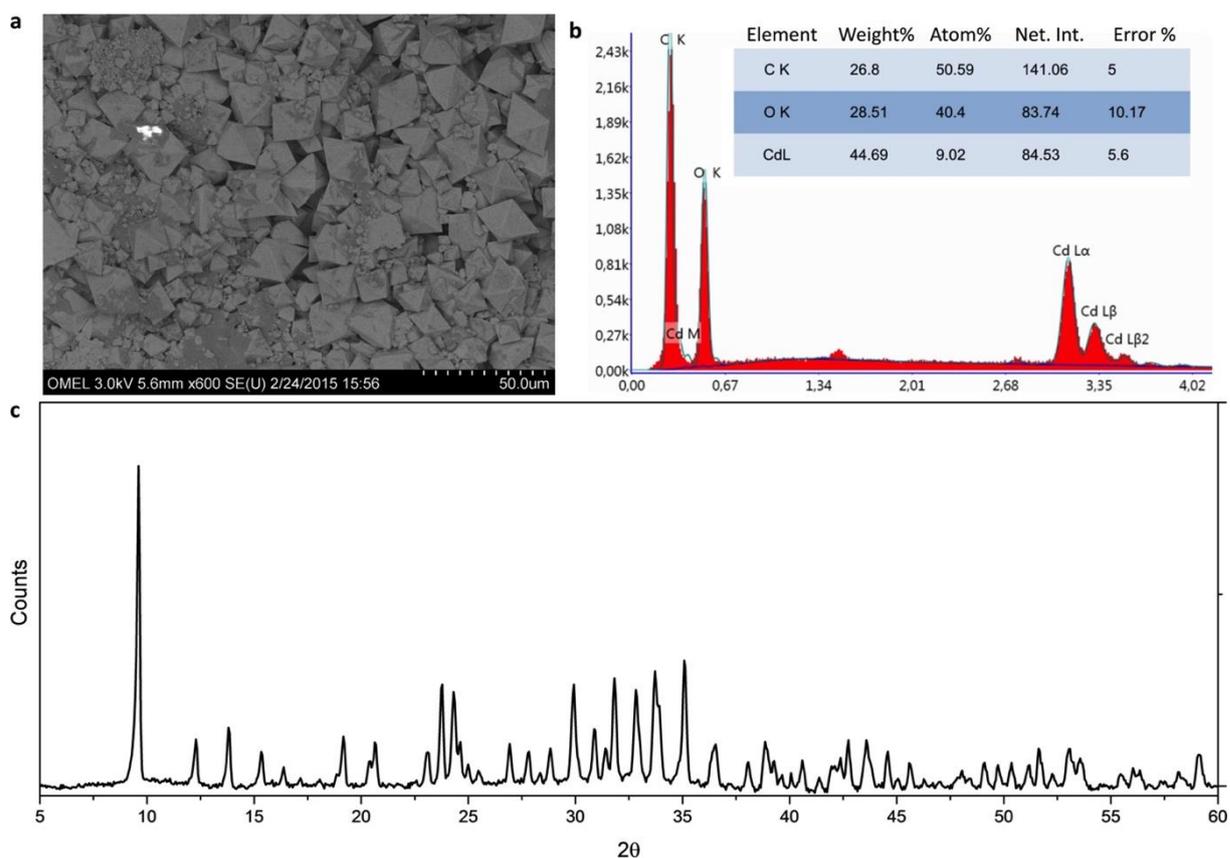

**Figure S3 | Analysis of the product obtained with acetic anhydride in absence of Se. a**, SEM micrographs revealed that large crystals were formed. **b**, The chemical composition was found to be very close to pure Cd(OAc)$_2$ by EDS analysis (obtained at 10 kV, x-axis in keV and y-axis in counts). The slightly higher carbon content possibly stems from oleate impurities. **c**, The corresponding powder XRD exhibited very sharp peaks. This phase could not be indexed with reasonable fit factors.

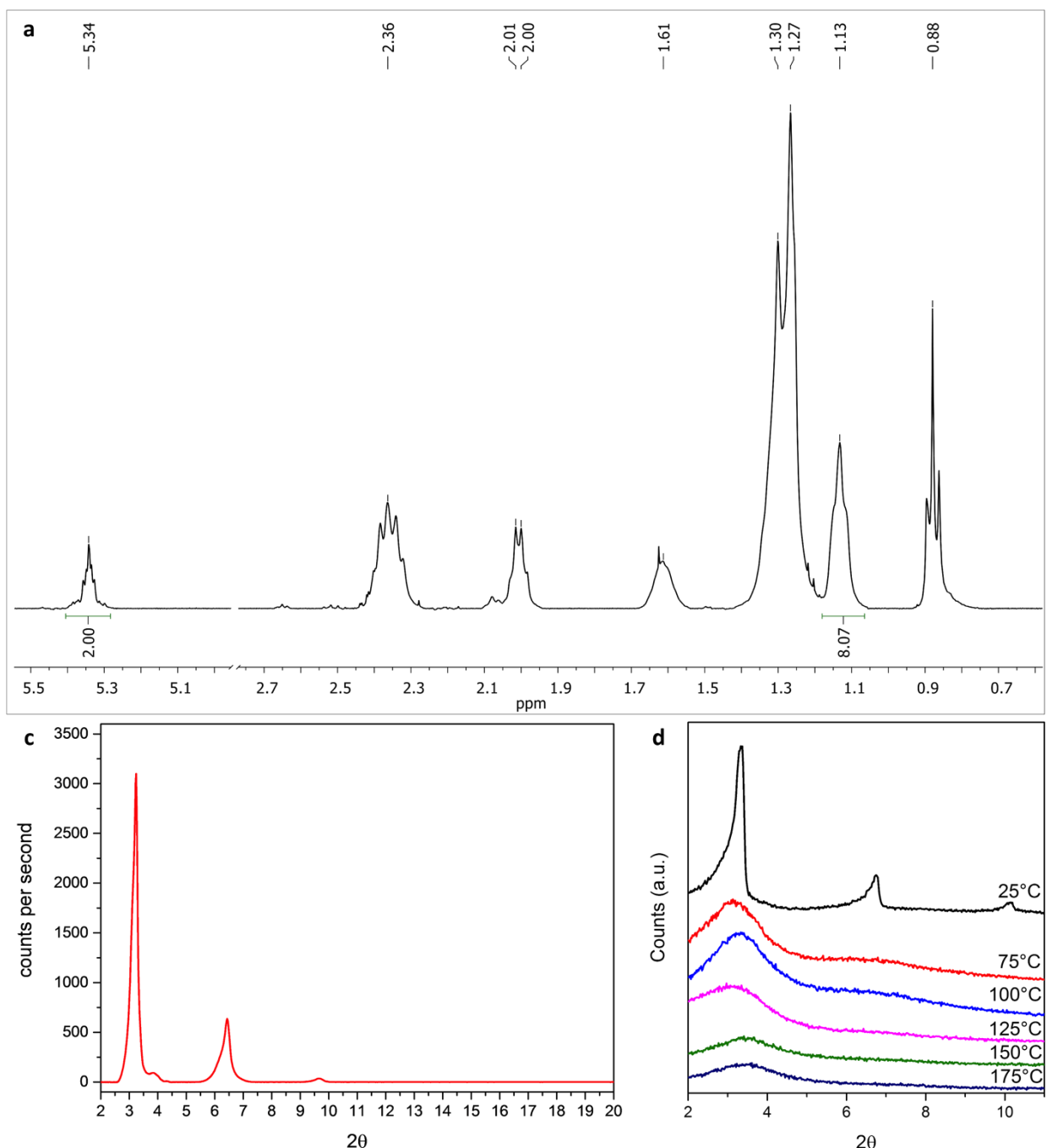

**Figure S4 | Analysis of the product obtained with propionic acid in absence of Se.**
**a**, ¹H-NMR spectrum (400 MHz, CDCl₃). The integral ratio of the olefinic (oleate moiety: m, δ = 5.34 ppm) and the methylic protons (of the propionate CH₃ terminus, t, δ = 1.13 ppm) suggested a composition close to Cd(oleate)$_{0.75}$(propionate)$_{1.25}$. However, diffusion ordered ¹H-NMR spectroscopy (inset) revealed that a mixture of Cd(oleate)$_1$(propionate)$_1$ and a minor fraction of Cd(propionate)$_2$ was present. **b**, The corresponding XRD pattern revealed the layered structure of this material, exhibiting clear [001], [002] and [003] reflections. **c**, When heated in a nitrogen atmosphere, the layered structure is lost already at relatively low temperatures.

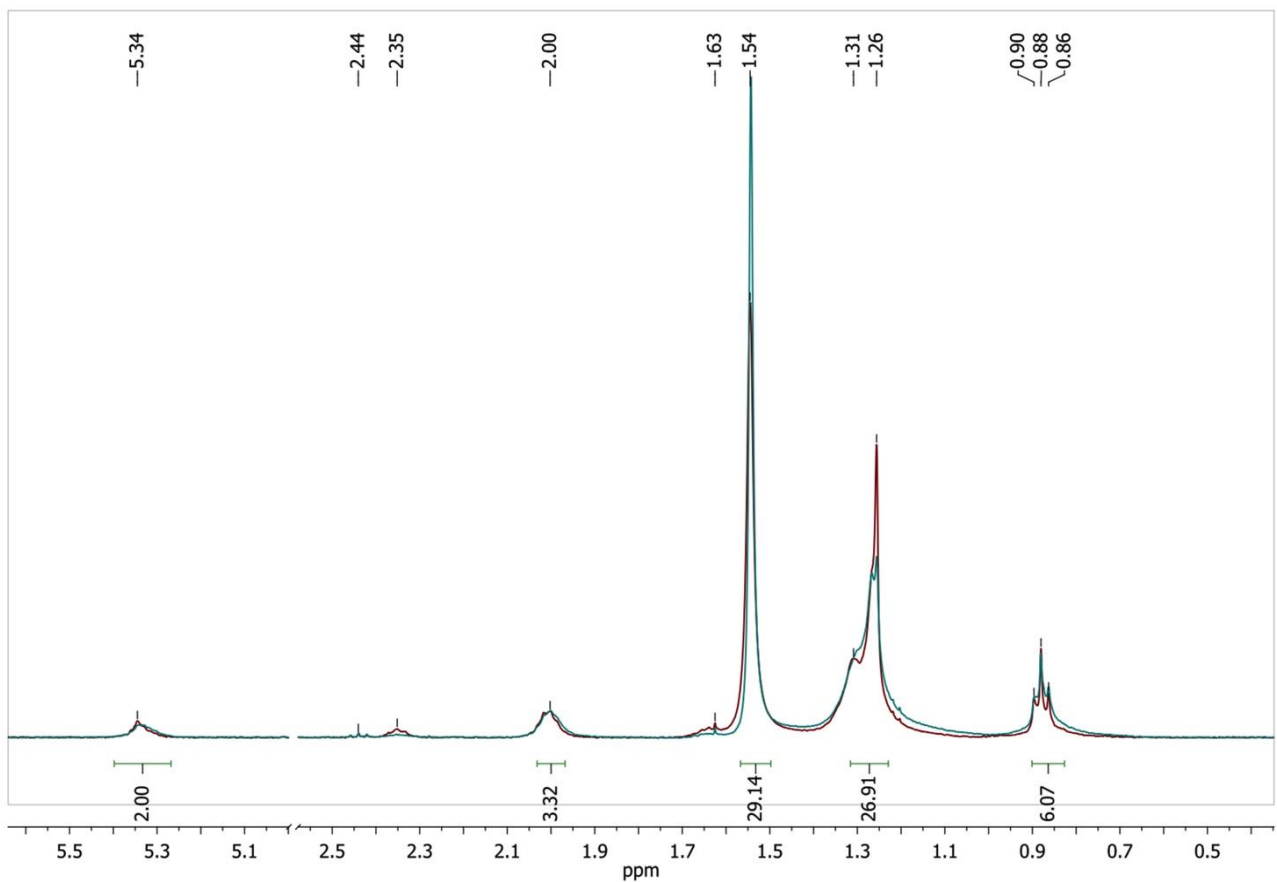

**Figure S5 | Analysis of the product obtained with butyric acid in absence of Se**. $^1$H-NMR (400 MHz) spectra of the product obtained with butyric acid (red curve) and pure Cd(oleate)$_2$ (blue curve) showed that oleate was not efficiently exchanged for butyrate since no clear signal stemming from butyrate protons could be detected.

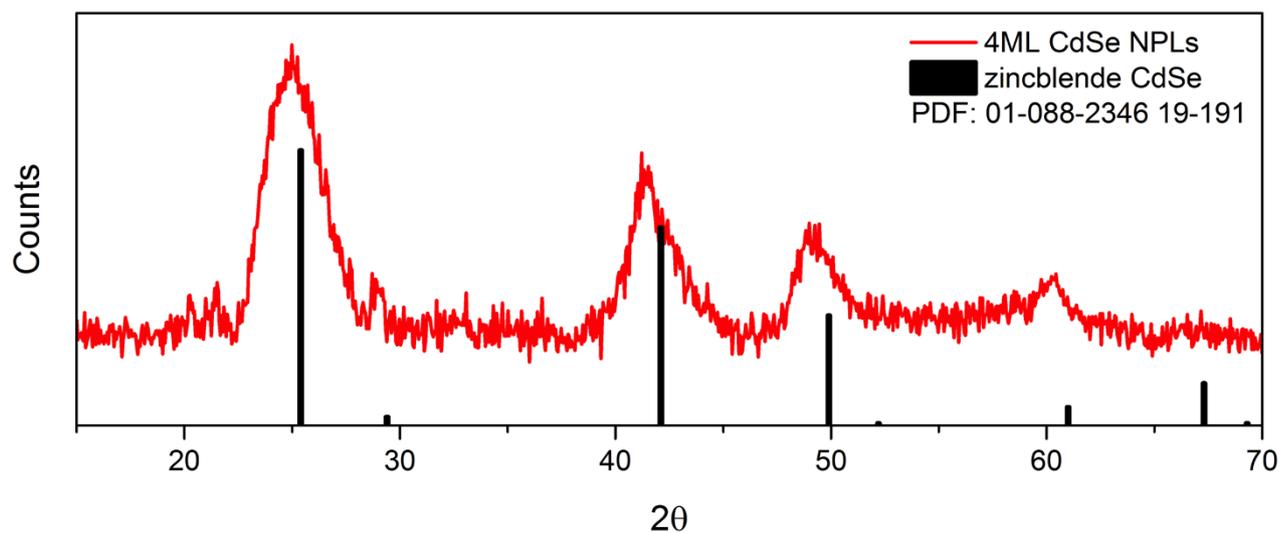

**Figure S6 | XRD analysis of 4-monolayer NPLs from a typical solvent-free synthesis.** The obtained diffraction pattern could be indexed to zincblende CdSe with small shifts towards lower angles, indicating a slightly larger unit cell.

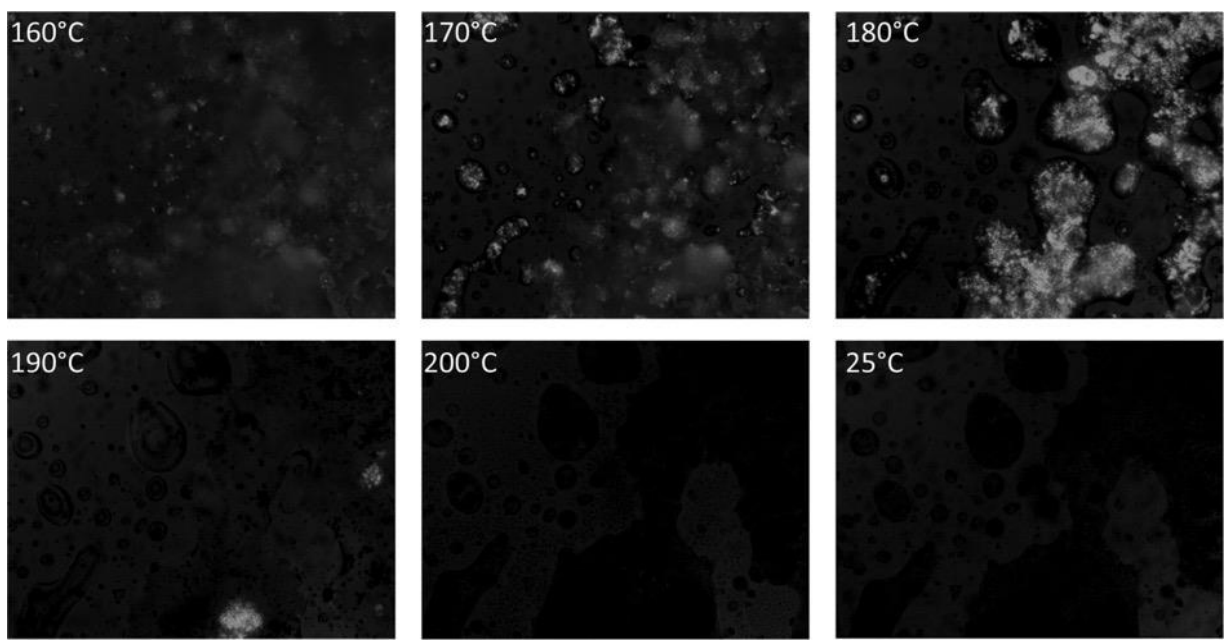

**Figure S7 | Polarized optical microscopy of Cd(propionate)$_2$/Se melts.** A blend of Cd(propionate)$_2$ and Se powders (3:1 molar ratio) was heated at a rate of 10 K/min (25-200 °C) under nitrogen atmosphere and images were taken at different temperatures. From room temperature up to 160 °C, the sample was optically thick and very little light intensity stemming from the birefringence of the crystalline powders was observed. More signal could be detected when the material started to melt (around 170-190 °C). The birefringence detected at this point was attributed to small crystalline fragments floating in the melt of Cd(propionate)$_2$. Once a homogeneous melt was obtained (200 °C), no birefringence was observed, which was also the case after the sample was cooled back to room temperature.

S25

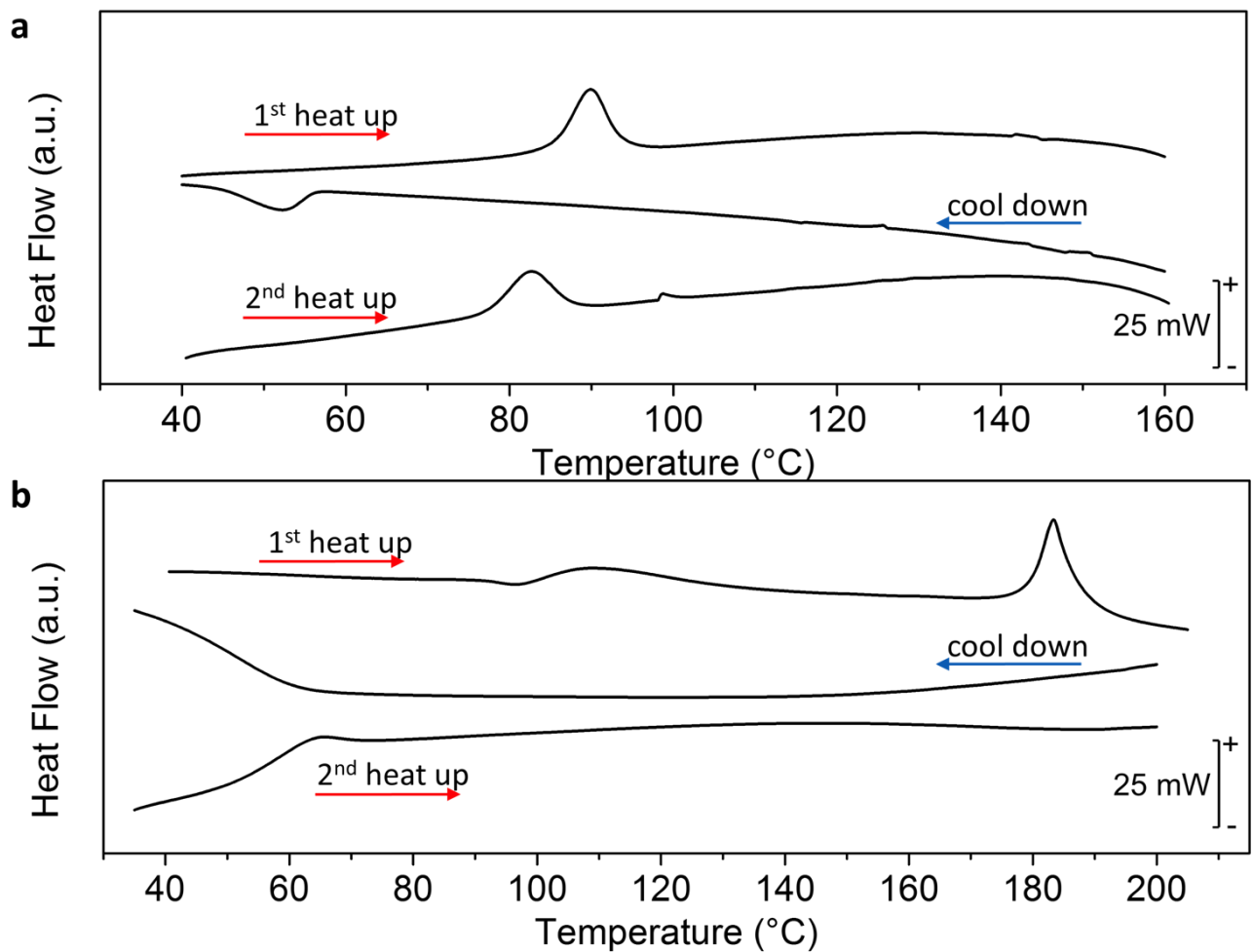

**Figure S8 | DSC of Cd(myristate)$_2$ and Cd(propionate)$_2$ recorded at a heating/cooling rate of 10 K/min under nitrogen atmosphere**. **a**, Cd(myristate)$_2$ melted at 87 °C and recrystallizes at 52 °C upon cooling. The melting temperature in the second heating step decreased to 82 °C. In contrast to Cd(propionate)$_2$, melting and crystallization are reversible for Cd(myristate)$_2$. **b**, Cd(propionate)$_2$ underwent a polymorphic transition at 110 °C and melted at 180 °C. The system did not recrystallize when cooled down, and amorphous Cd(propionate)$_2$ was obtained.

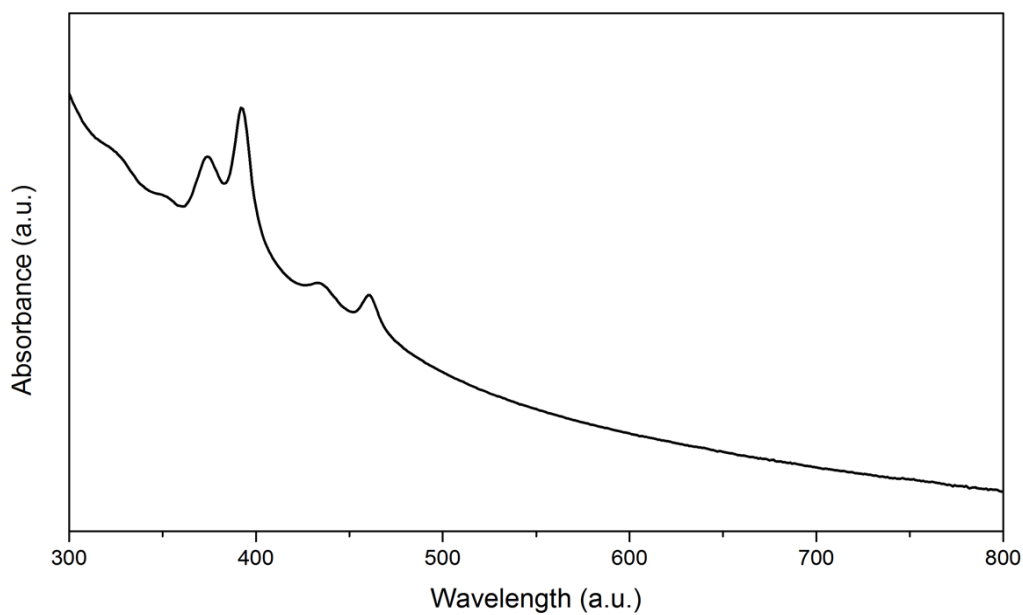

**Figure S9 | Absorption spectrum of NPLs synthesized with amorphous Cd(propionate)₂.** Amorphous Cd(propionate)$_2$ was obtained by thermal pre-treatment (see main text and Figure S8), mixed with elemental Se and reacted at 200 °C for 4 h. Absorption features of 3-monolayer (372 nm and 393 nm) and 4-monolayer (435 nm and 461 nm) zincblende CdSe NPLs can be observed. The fact that NPLs could form from amorphous precursors proved conclusively that NPLs form in a completely isotropic environment.

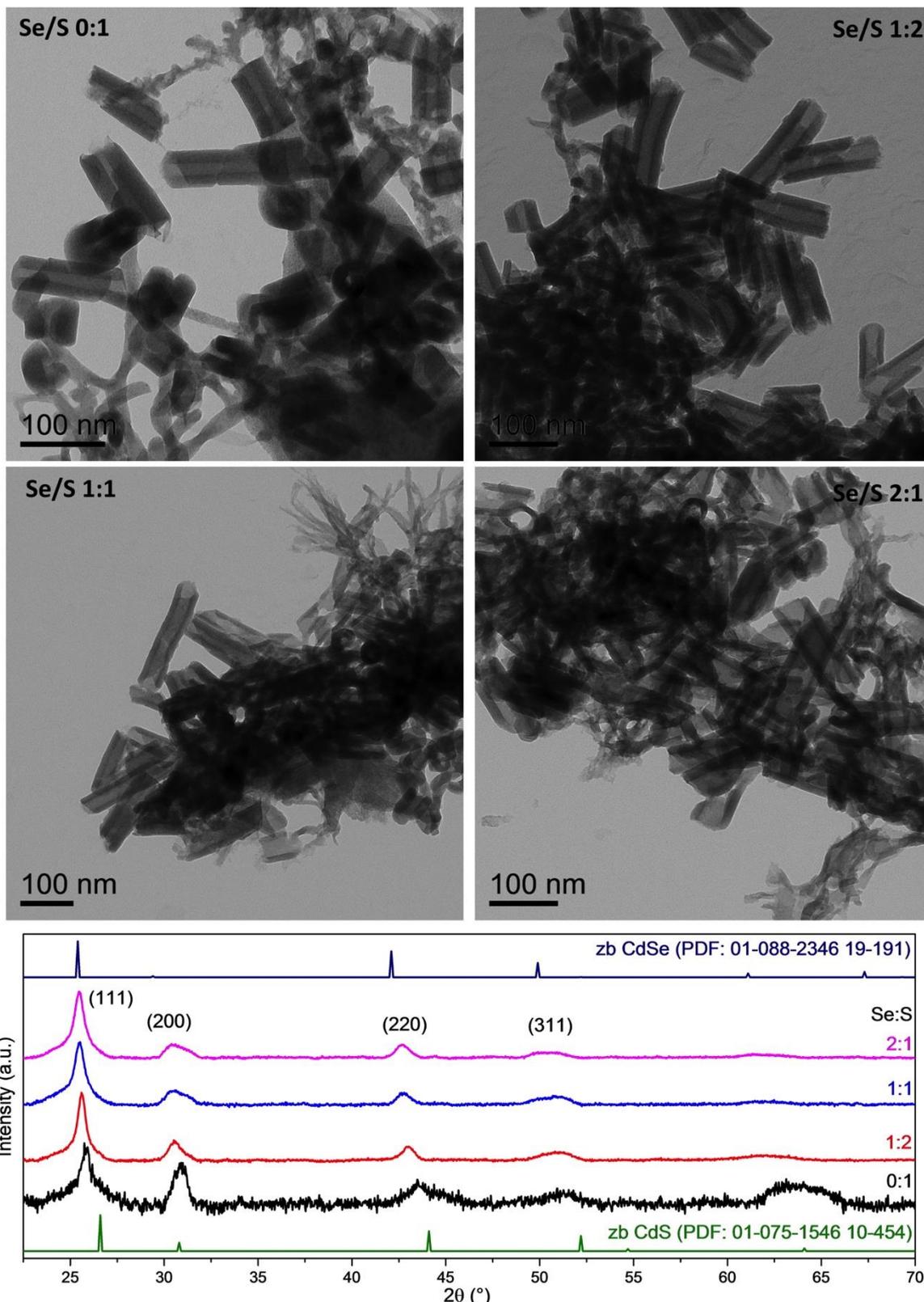

**Figure S10 | TEM images and corresponding XRD patterns of samples obtained from different molar ratios Se:S.** The molar ratio of Cd:(S+Se) was 3:1 in all cases. Blended powders were heated for 18 h at 200 °C and dispersed and washed as described above. All samples tended to roll up, most likely due to lattice strain. The XRD analysis revealed the zincblende crystal structure and the alloyed NPLs of the type $CdS_{1-x}Se_x$ since all reflections are progressively shifted to lower angles with increasing Se content.

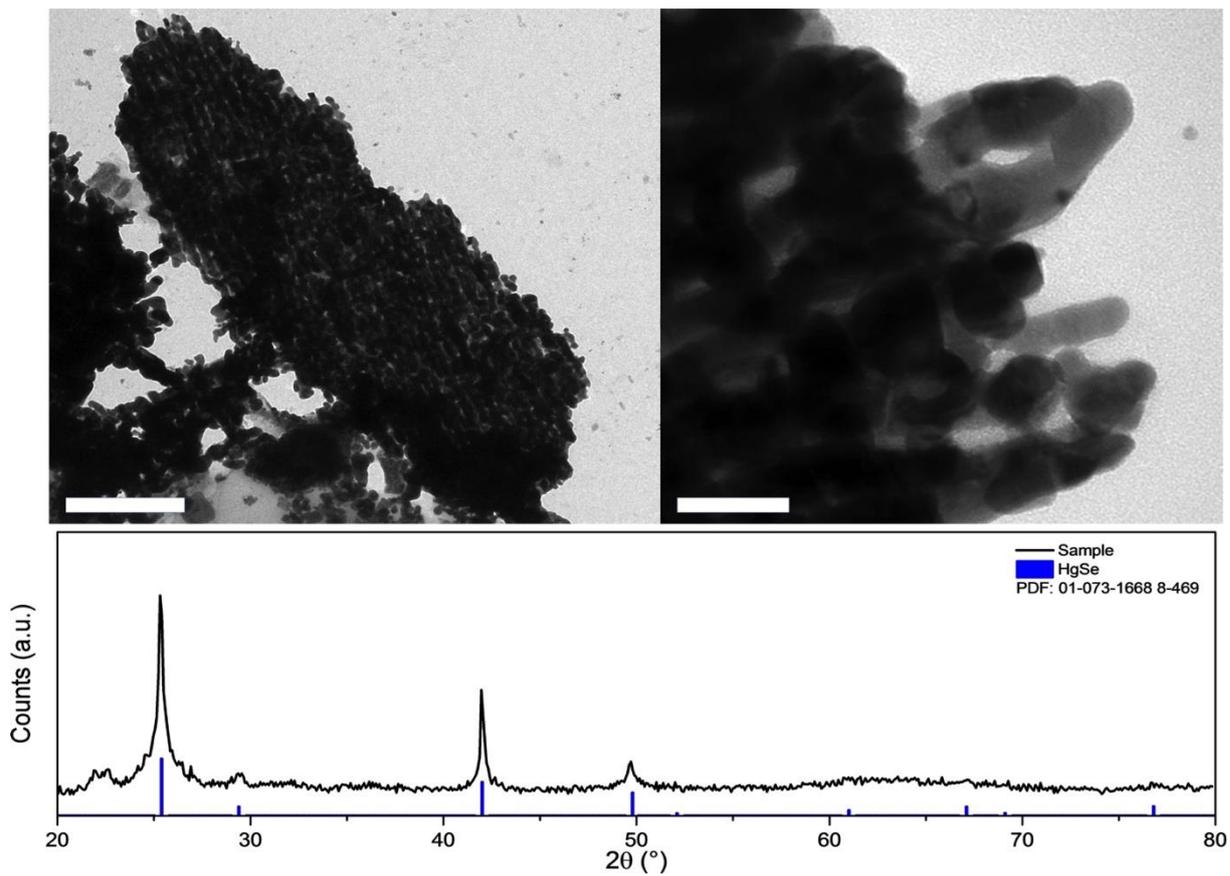

**Figure S11 | TEM micrographs and XRD pattern of HgSe produced in melts of Hg(OAc)$_2$ and elemental Se.** No atomically flat HgSe NPLs were observed. However, the crystals still exhibited an elongated, non-cubic habit. Scale bars correspond to 200 nm (bottom left) and 50 nm (bottom right). The XRD pattern could be indexed to zincblende HgSe (Tiemannite). Experiments with other Hg(carboxylate)$_2$ precursors led to similar results.

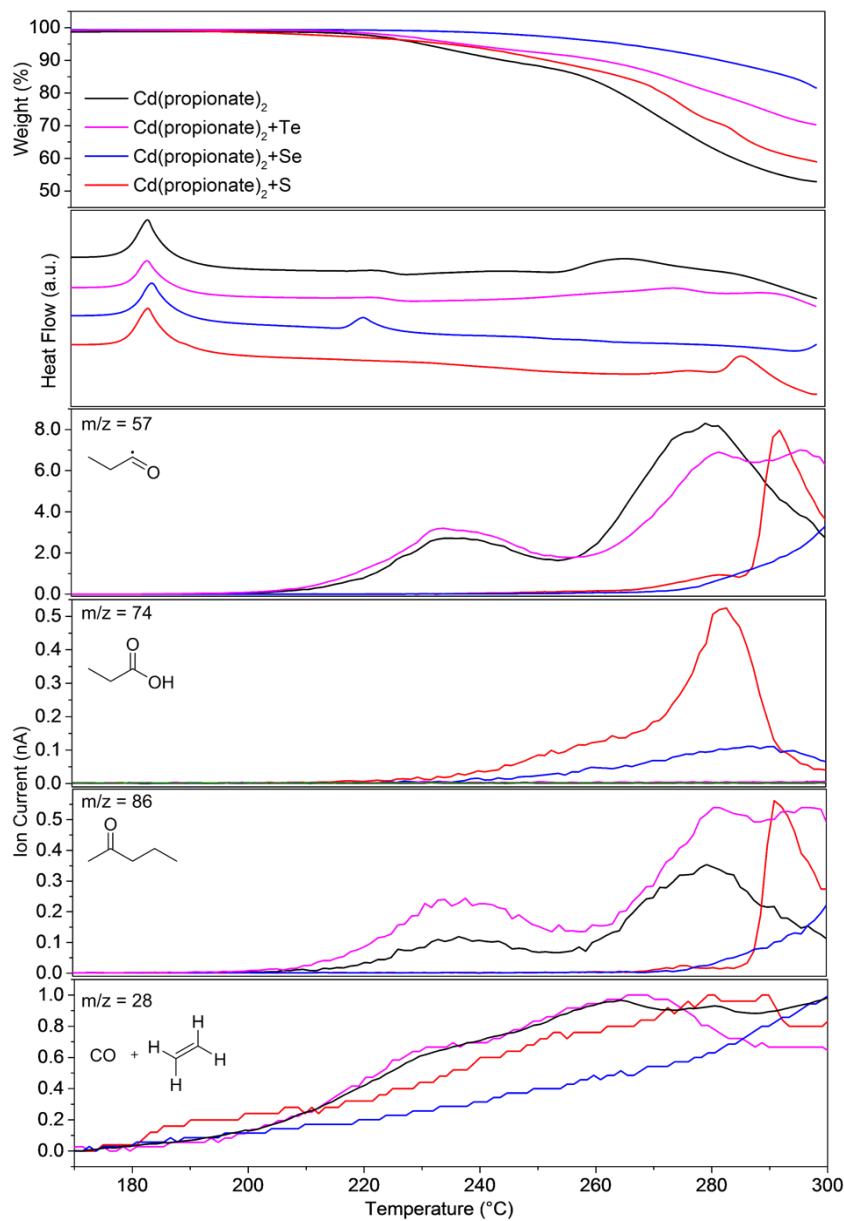

**Figure S12 | TGA/DSC data with additional mass traces of Cd(propionate)$_2$ alone or with Te, Se, or S.** Thermal decomposition of Cd(propionate)$_2$ starts around 200 °C. In the relevant temperature range (180-240 °C) the typical propionate decomposition products (propionyl radical and 2-pentanone) were observed only in samples that did not contain Se or S. This finding led to the proposed mechanism depicted in Fig. 3e in the main text where elemental Se (S) reacts with the propionyl radicals to form dipropionyl selenide (sulfide), which reacts with remaining Cd(propionate)$_2$ to yield CdSe under release of propionyl cations (from dipropionyl selenide) and propionate ions [from Cd(propionate)$_2$]. While propionate ions can bind to the surface of formed CdSe NPLs, propionyl cations are unstable and decompose to ethene, carbon monoxide, and protons. Indeed, mass traces for m/z=28 (which corresponds to both CO and ethene) showed increasing ion currents for all samples (starting already at 180-200 °C), while traces for m/z=74 (corresponds to propionic acid) showed significant increase only for samples containing Se or S. This indicates that with increasing CdSe formation, more and more propionate ions desorb from the crystal surfaces and capture some of the released protons stemming from the decomposition of propionyl cations forming propionic acid.

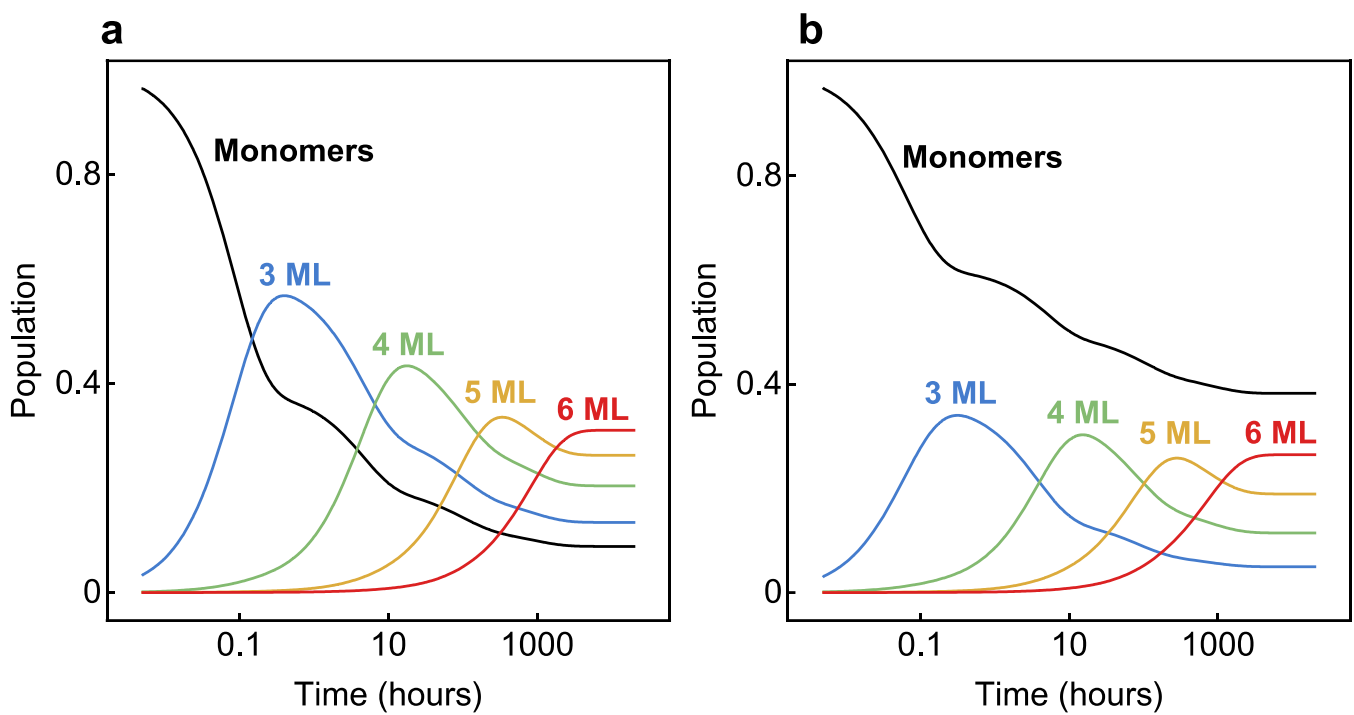

**Figure S13 | Comparison of simplified model based on first-order kinetics and more realistic model including geometric considerations. a**, Solution of the simple model at T = 200 ˚C with $k_0$ fitted to experimentally observed timescales (see main text). **b**, More realistic model solved with the value for $k_0$ extracted from the first-order model.

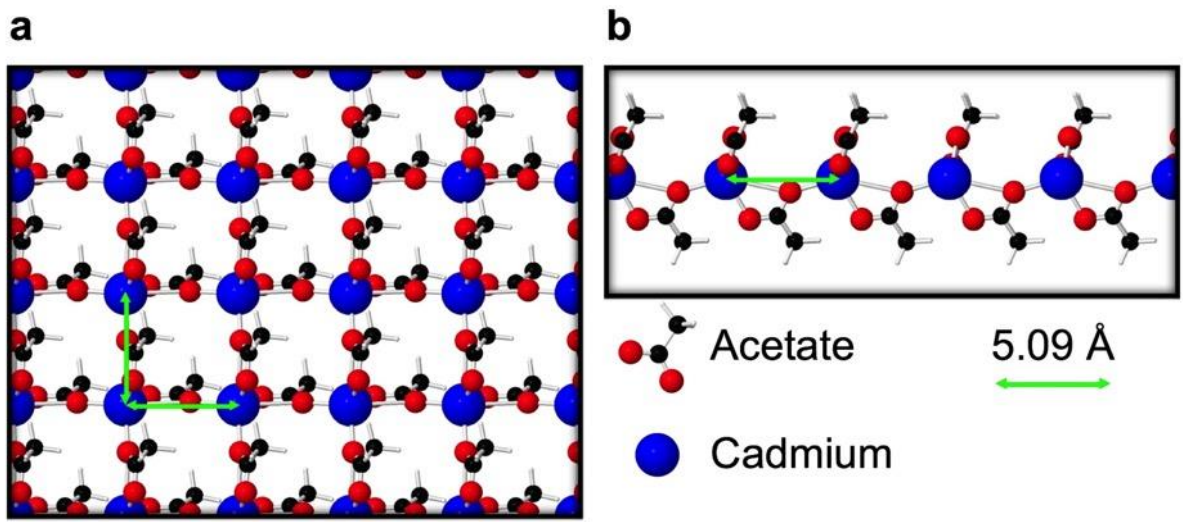

**Figure S14 | Structure of the Cd-acetate phase used for our DFT calculations.** We used 2-dimensional coordination polymers of Cd(OAc)$_2$ to estimate the energy of the Cd-acetate precursor. Projections of this 2-dimensional polymer from the top and the side are shown in **a** and **b**, respectively.

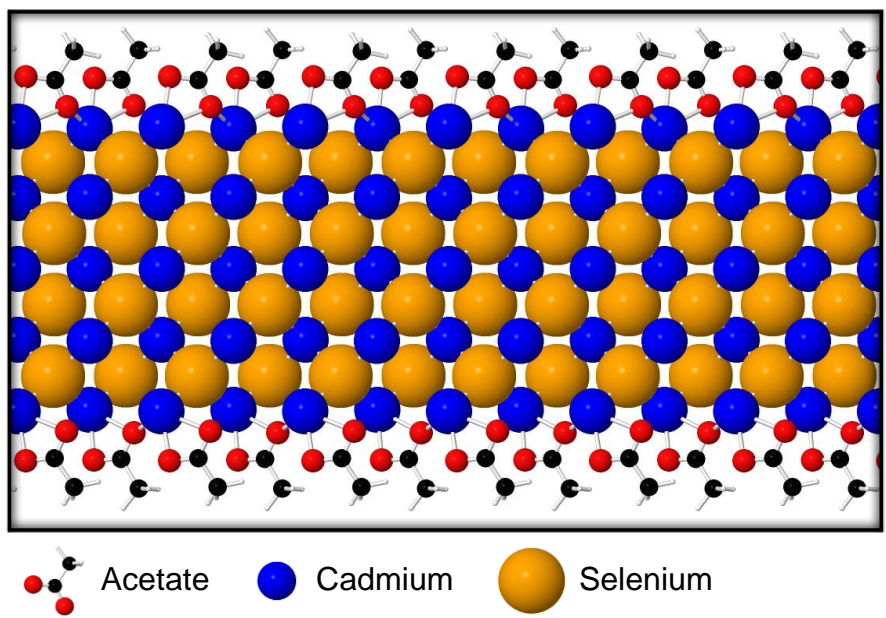

**Figure S15 | Side view of a 4-monolayer thick CdSe(001) zincblende nanoplatelet.** The {001} surfaces are Cd-terminated and passivated by acetate molecules that form bidentate bonds to the Cd atoms on the surface.

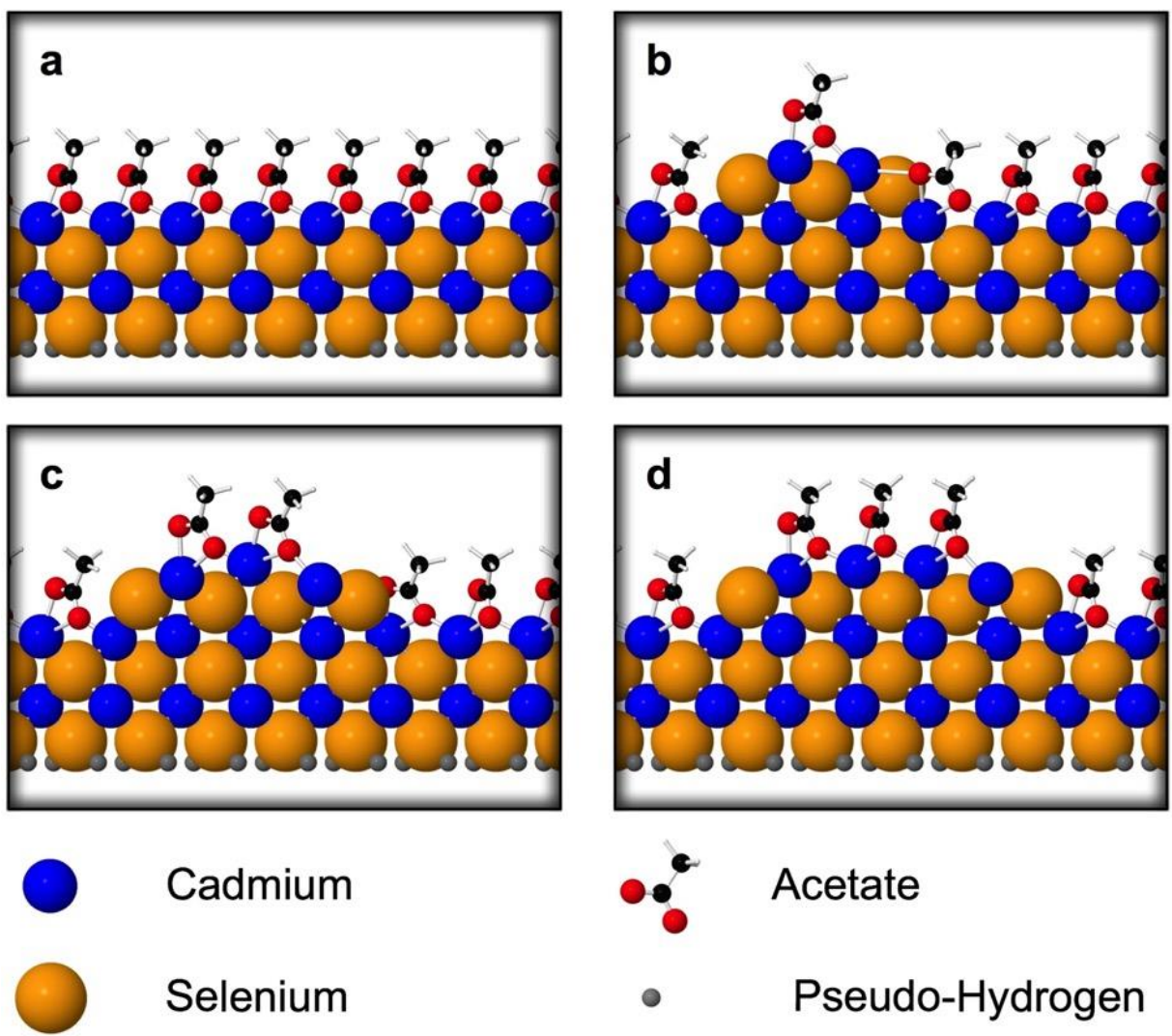

**Figure S16 | Calculation of the edge formation energy. a**, The reference geometry is a two-monolayer-thick CdSe(001) slab on top passivated by acetates and on the bottom by pseudo-hydrogen atoms. **b-d**, Different computational supercells with electronic closed-shell structures following the electron counting rule. We obtained formation energies of 0.58 eV ($E_L = 40$ meV/Å), 0.42 eV ($E_L = 24$ meV/Å), and 0.64 eV ($E_L = 47$ meV/Å) for **b**, **c**, and **d**, respectively.

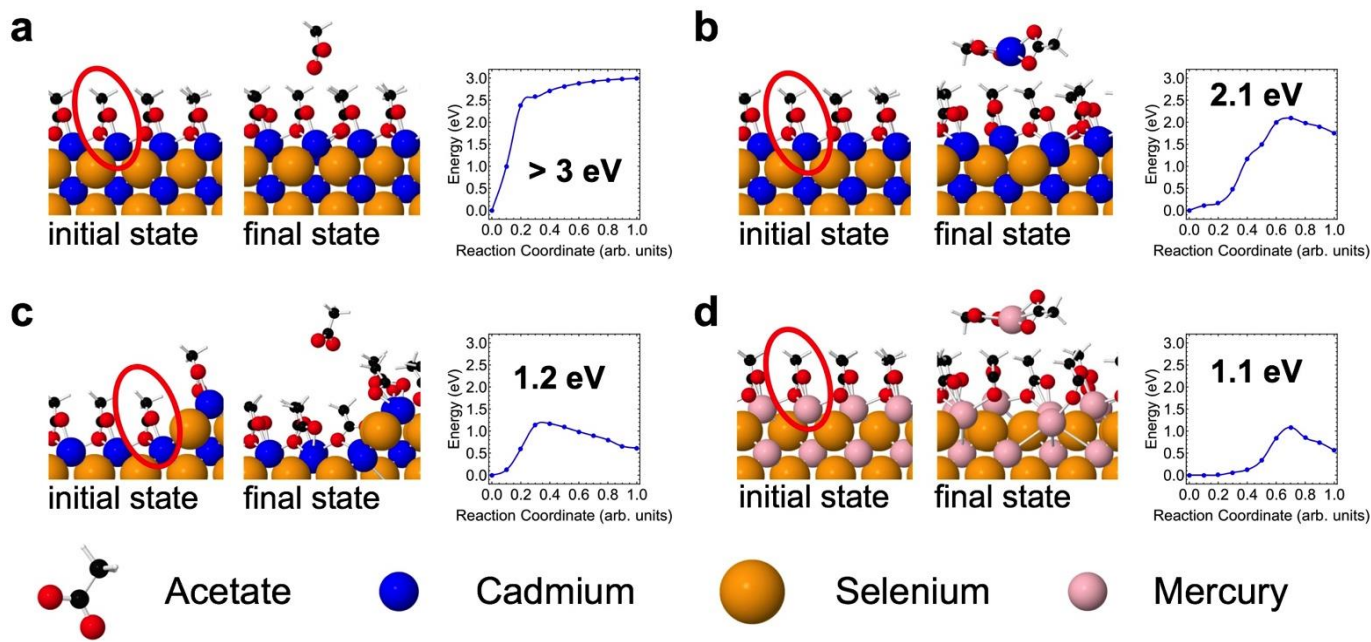

**Figure S17 | Calculation of the minimum energy path for selected surfactant desorption reactions.** Desorbing molecules are marked with a red circle in the initial state. **a**, Desorption of acetate from a perfectly flat CdSe(001) surface has an energy barrier of more than 3 eV. **b**, Desorption of Cd-acetate from this surface has a reduced barrier of 2.1 eV. **c**, The barrier for desorption of the acetate next to a surface step is only 1.2 eV. **d**, On flat and passivated HgSe(001) surface the binding energy of Hg-acetate molecule is 1.1 eV.

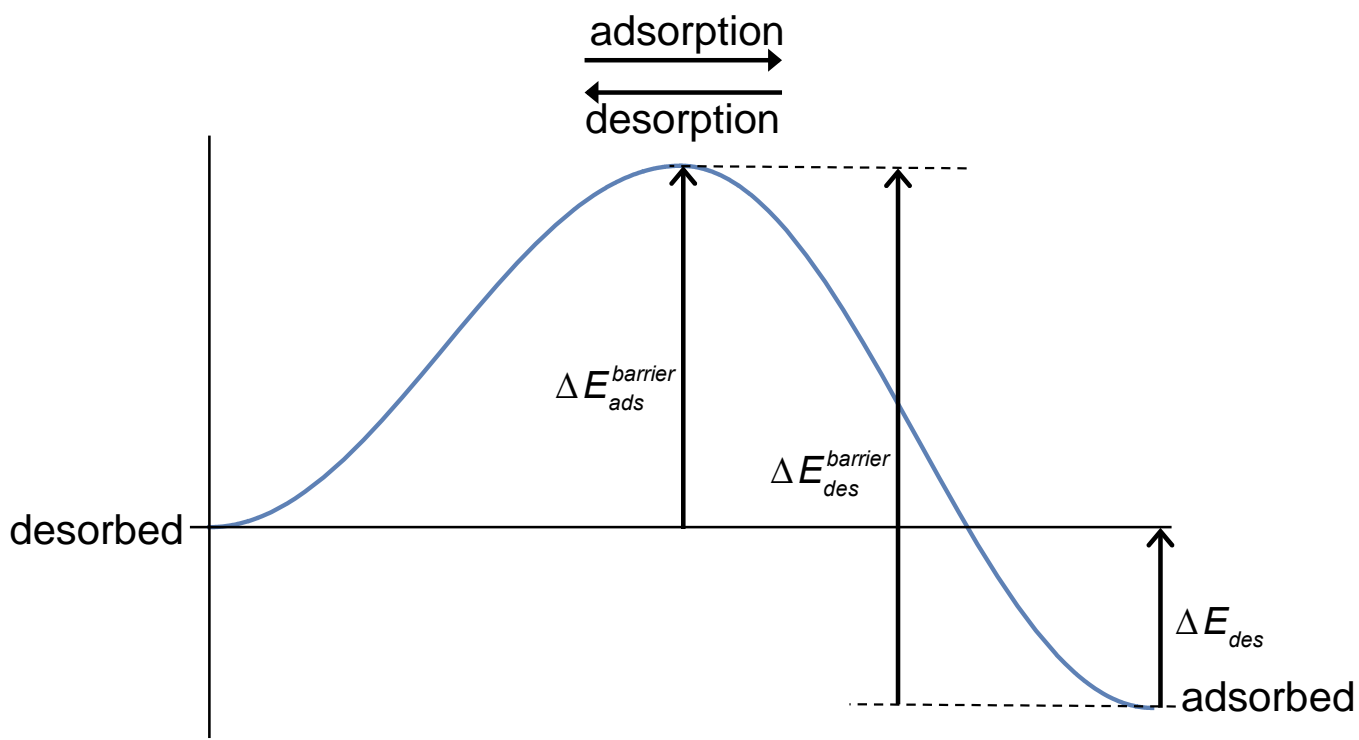

**Figure S18 | A schematic potential-energy surface for monomer adsorption and desorption reactions**. The reaction coordinate going from left to right corresponds to the reaction of transferring a monomer from the melt to a crystal surface. The corresponding barrier is known from DFT calculations of surfactant desorption reactions. The reaction coordinate going from right to left is the desorption reaction for which the barrier is given by the sum of the monomer adsorption barrier and the energy difference between desorbed and adsorbed state.

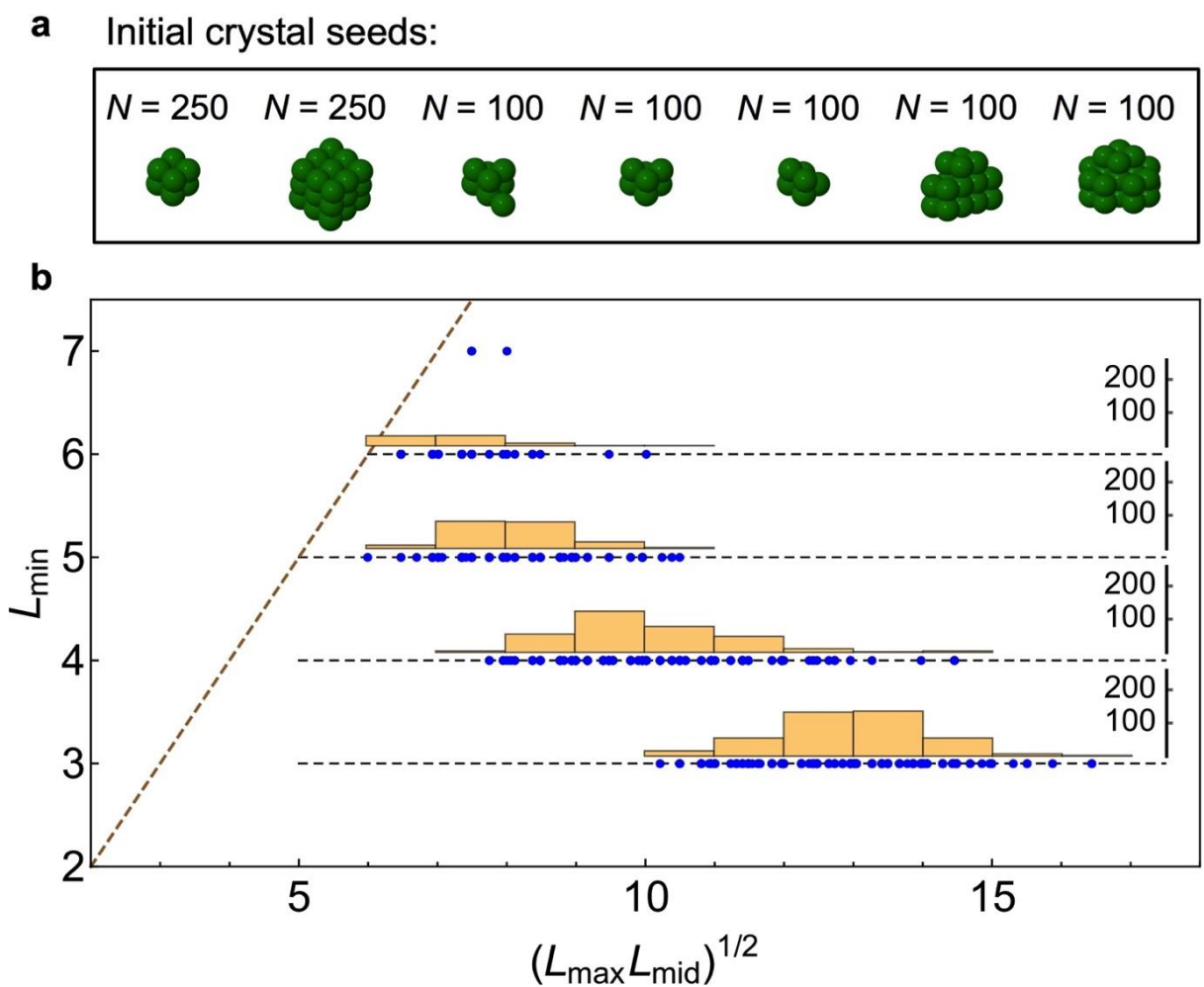

**Figure S19 | Shape distribution of nanocrystallites grown using kinetic Monte Carlo simulations**. **a,** Crystallites were started as small crystal seeds and then grown at 1000 K. For each of the shown crystal seeds we started $N$ simulation runs, as given in the figure. **b,** The grown crystallites are each characterized here by the shape of a rectangular box (in units of monomers) closely corresponding to the crystallite shape. The plot abscissa is a linear dimension given by the geometric average of the two largest box dimensions, and the ordinate is the shortest linear dimension of the box. Each simulation outcome is plotted as a blue dot and the histograms show the size distributions of platelets with thicknesses ranging from 3 to 6 monolayers. The diagonal dashed line depicts the (hypothetical) case of isotropic shapes, i.e. cubical nanocrystallites. Dotted lines indicate crystallites with various nanoplatelet shapes: that is, thicknesses from three to six monolayers and large areas. The crystallites of three and four monolayers tend to have the largest area and strongly resemble our experimentally synthesized nanoplatelets. For thicker crystallites a trend toward forming more isotropic shapes is apparent; this trend is also consistent with our experimental results.

## S6. Supplementary references

S1. Mahler, B., Nadal, B., Bouet, C., Patriarche, G. & Dubertret, B. Core/shell colloidal semiconductor nanoplatelets. *J. Am. Chem. Soc.* **134**, 18591-18598 (2012).

S2. Kuvadia, Z. B. & Doherty, M. F. Spiral growth model for faceted crystals of non-centrosymmetric organic molecules grown from solution. *Cryst. Growth Des.* **11**, 2780-2802 (2011).

S3. Lovette, M. A. & Doherty, M. F. Predictive modeling of supersaturation-dependent crystal shapes. *Cryst. Growth Des.* **12**, 656-669 (2012).

S4. Kim, S. H., Dandekar, P., Lovette, M. A. & Doherty, M. F. Kink rate model for the general case of organic molecular crystals. *Cryst. Growth Des.* **14**, 2460-2467 (2014).

S5. Ohara, M. & Reid, R. C., *Modeling Crystal Growth Rates from Solution*. (Prentice-Hall, 1973, 1973).

S6. Perdew, J. P., Burke, K. & Ernzerhof, M. Generalized gradient approximation made simple. *Phys. Rev. Lett.* **77**, 3865-3868 (1996).

S7. Perdew, J. P., Burke, K. & Ernzerhof, M. Generalized gradient approximation made simple [Phys. Rev. Lett. 77, 3865 (1996)]. *Phys. Rev. Lett.* **78**, 1396-1396 (1997).

S8. Kresse, G. & Hafner, J. Ab initio molecular dynamics for liquid metals. *Phys. Rev. B* **47**, 558-561 (1993).

S9. Kresse, G. & Furthmüller, J. Efficient iterative schemes for ab initio total-energy calculations using a plane-wave basis set. *Phys. Rev. B* **54**, 11169-11186 (1996).

S10. Van de Walle, C. G., Laks, D. B., Neumark, G. F. & Pantelides, S. T. First-principles calculations of solubilities and doping limits: Li, Na, and N in ZnSe. *Phys. Rev. B* **47**, 9425-9434 (1993).

S11. Van de Walle, C. G. & Neugebauer, J. First-principles calculations for defects and impurities: Applications to III-nitrides. *J. Appl. Phys.* **95**, 3851-3879 (2004).

S12. Stephan, L. & Alex, Z. Accurate prediction of defect properties in density functional supercell calculations. *Model. Simul. Mater. Sci. Eng.* **17**, 084002 (2009).

S13. Henkelman, G., Uberuaga, B. P. & Jónsson, H. A climbing image nudged elastic band method for finding saddle points and minimum energy paths. *J. Chem. Phys.* **113**, 9901-9904 (2000).

S14. Fichthorn, K. A. & Weinberg, W. H. Theoretical foundations of dynamical Monte Carlo simulations. *J. Chem. Phys.* **95**, 1090-1096 (1991).

S15. Ott, F. D., Spiegel, L. L., Norris, D. J. & Erwin, S. C. Microscopic Theory of Cation Exchange in CdSe Nanocrystals. *Phys. Rev. Lett.* **113**, 156803 (2014).